\documentclass[elsart12,eqsecnum,graphics,cite,nofootinbib]
{revtex4-2}

\usepackage{slashed}
\usepackage{amsmath}
\usepackage{amssymb}
\usepackage{amsfonts}
\usepackage{array}
\usepackage{graphicx}
\usepackage{subcaption}
\usepackage{mathrsfs}
\usepackage{multirow}
\usepackage{siunitx}
\usepackage{float}
\usepackage{hyperref}
\hypersetup{
    colorlinks,
    citecolor=blue,
    filecolor=blue,
    linkcolor=blue,
    urlcolor=blue
}
\usepackage{bm}
\usepackage{xcolor}
\usepackage{academicons}
\usepackage{titlesec}
\titlespacing*{\section}{0pt}{20pt}{10pt}
\titlespacing*{\subsection}{0pt}{10pt}{10pt}
\renewcommand{\v}[1]{ \ensuremath{ {\bm{#1}} }}

\begin{document}


\title{
Entanglement Enabled Intensity Interferometry \\
in ultrarelativistic ultraperipheral nuclear collisions 
}


\author{James Daniel Brandenburg}
\email{brandenburg.89@osu.edu}
\affiliation{Department of Physics, The Ohio State University, Columbus, OH 43210, USA}

\author{Haowu Duan}
\email{haowu.duan@uconn.edu}
\affiliation{Physics Department, University of Connecticut, 2152 Hillside Road, Storrs, CT 06269, USA}

\author{Zhoudunming Tu}
\email{zhoudunming@bnl.gov}
\affiliation{Physics Department, Brookhaven National Laboratory, Upton, NY 11973, USA}
\author{Raju Venugopalan}
\email{raju.venugopalan@gmail.com}
\affiliation{
Physics Department, Brookhaven National Laboratory, Upton, NY 11973, USA\\
Center for Frontiers in Nuclear Science, Stony Brook University, NY 11794, USA}
\author{Zhangbu Xu}
\email{zxu22@kent.edu}
\affiliation{Department of Physics, Kent State University, Kent, OH 44242, USA }

\date{\today}

\begin{abstract}

An important tool in studying the sub-femtoscale spacetime structure of matter in ultrarelativistic heavy-ion collisions is Hanbury-Brown-Twiss (HBT) intensity interferometry of identical particles in the final state of the  collisions. We propose that a variant of the entanglement enabled intensity interferometry ($E^2 I^2$) framework introduced by Cotler and Wilczek can provide a powerful alternative to HBT interferometry in extracting fundamental nonperturbative features of QCD at high energies. We apply this framework to demonstrate that the spatial distributions of color singlet (pomeron) configurations in nuclei are sensitive to measurements of exclusive resonant decays of $\rho$-mesons into $\pi^\pm$-pairs in ultrarelativistic ultraperipheral nuclear collisions (UPCs) at RHIC and the LHC. A preliminary analysis suggests that the model-independent extraction of pomeron distributions will require careful treatment of the interplay of $E^2 I^2$ in the vector meson exclusive decay with the incoherent cross-section for exclusive vector meson production. The $E^2 I^2$ framework developed here is quite general. It can also be employed as a tool to extract information on the spin structure of pomeron couplings as well as enhance the discovery potential for rare odderon configurations from exclusive vector meson decays into few-particle final states both in UPCs and at the Electron-Ion Collider.

\end{abstract}

\maketitle

\section{Introduction}

High-energy heavy-ion collisions recreate on earth the hot and dense quark-gluon plasma (QGP), which is a non-Abelian fluid that only existed previously in the very early universe~\cite{Sun:2022xjr,Busza:2018rrf}. An essential tool in characterizing the spacetime structure of the QGP~\cite{Goldhaber:1959mj,Lisa:2005dd,Harris:2023tti} is 
Hanbury-Brown--Twiss (HBT) intensity interferometry, originally developed for astronomical imaging~\cite{brown1956twiss,brown_correlation_1956}. This technique determines the spacetime structure of a chaotic source from measurements in separate detectors of correlations in the field intensities of identical particles emitted by the source~\cite{Baym:1997ce}. Intensity interferometry can be understood as a generic wave interference phenomenon and has been demonstrated with both classical and nonclassical light~\cite{doi:10.1126/sciadv.adh1439}.  
It is an essential tool across various fields of physics, from astronomy~\cite{doi:10.1126/science.1208192}, high- and low-energy nuclear physics~\cite{Wiedemann:1999qn,Boal:1990yh}, and quantum optics~\cite{chrapkiewicz_hologram_2016}--see \cite{aspect2020hanburry} for a review. All these cases require the intensity interferometry to be performed on measurements of indistinguishable particles.

A novel idea proposed by Cotler and Wilczek~\cite{cotler_entanglement_2015} and developed further by Cotler, Wilczek and Borish~\cite{COTLER2021168346}, suggested that intensity interference effects could be recovered in measurements of nonidentical particles by passing the emitted particles through a device that performs a unitary transformation entangling their wavefunctions, and subsequently filters the entangled state prior to their measurement in a detector. They dubbed this phenomenon ``entanglement enabled intensity interferometry" ($E^2 I^2$). Cotler, Wilczek and collaborators later demonstrated this $E^2 I^2$ effect in optical intensity interferometry experiments utilizing distinguishable photons of two distinct wavelengths~\cite{qu_chromatic_2020,liu_improved_2021}. 
 
The STAR collaboration has speculated that this 
$E^2 I^2$ phenomenon may provide a possible explanation for patterns observed in exclusive decays of vector mesons ($\rho\rightarrow \pi^+\pi^-, \phi\rightarrow K^+K^-, J/\psi\rightarrow e^+ e^-,\cdots$) measured in ultrarelativistic ultraperipheral nuclear collisions (UPCs) at the Relativistic Heavy Ion Collider (RHIC)~\cite{STAR:2022wfe}.  In UPCs at RHIC, and at the Large Hadron Collider (LHC), beams of heavy ultrarelativistic nuclei generate extremely powerful electromagnetic fields that can lead to the formation of electromagnetic pairs and the realization of striking QED phenomena such as the Breit-Wheeler effect~\cite{STAR:2019wlg,Brandenburg:2022tna}.  Further, UPCs allow for clean studies of strongly interacting matter because the equivalent Weizs\"{a}cker-Williams (WW) photons~\cite{PhysRev.45.729,weizsacker_ausstrahlung_1934} from one of the nuclear beams can scatter (directly) off quark and (indirectly) off gluon fields in nuclei, producing strongly interacting subatomic particles in the final state~\cite{Bertulani:2005ru,Klein:2017nqo}. 

A particularly powerful tool in UPC measurements are so-called exclusive diffractive photoproduction processes which provide clean information on the spacetime tomography of quarks and gluons in nuclei. In these processes, the decay products of a single final state (such as a vector meson or jet) can be studied in isolation separated by a gap in rapidity from the struck nucleus. 
Triggering on such events is further aided by selecting ultraperipheral collisions that undergo mutual Coulomb excitation, and subsequent dissociation, leading to the measurement of neutrons with beam-energy in zero degree calorimeters (ZDC). A broad range of experiments with exclusive resonant decays of vector mesons have now been performed at RHIC~\cite{STAR:2002caw,STAR:2011wtm,STAR:2007elq,STAR:2017enh,STAR:2019yox,STAR:2023gpk} and the LHC~\cite{ALICE:2021tyx,Goncalves:2020vdp}.

The pattern suggestive of $E^2 I^2$ observed by STAR in exclusive $\rho^0$-meson decays into $\pi^\pm$ pairs is a strong $\cos{2\phi}$ and a modest $\cos{4\phi}$ modulation in the coherent cross-section. The coherent cross-section corresponds to intact nuclei as determined by the ZDC measurements. The angle $\phi$ is the azimuthal angle between $\v q_\perp=(\v {p_1}+\v {p_2})$ and $\v P_\perp=(\v {p_1} - \v{p_2})$, with $\v {p_{1,2}}$ being the momentum vectors of the daughter $\pi^\pm$, projected along the plane orthogonal to the beam axes. The observed $\cos{2\phi}$ modulation is strongest at low transverse momentum, peaking with a value of $\sim40\%$ at a $|\v q_\perp|\approx 20$ MeV$/c$ and shows a wave interference structure exhibiting a minimum and second maximum around $|q_\perp|\approx 120$ MeV$/c$~\cite{STAR:2022wfe} -- the STAR plot demonstrating this is shown in Fig.\ref{fig:coherent+incoherent} -- later in the paper. The $\cos{4\phi}$ modulation is seen for  $|\v {q_\perp}|\approx120$ MeV$/c$. The $\cos{2\phi}$ modulation was recently confirmed by the ALICE collaboration at the LHC~\cite{ALICE_upc_spin_qm}.

The experimental results have been compared with several models that take into account the linear polarization of the WW photons initiating the scattering and the interference between the amplitudes corresponding to the likelihood that the $\rho$-meson is produced off one nucleus or the other~\cite{Goncalves:2017wgg,Bendova:2020hbb,Xing:2020hwh,Zha:2020cst,Hagiwara:2020juc,Guzey:2022qvc,Mantysaari:2022sux,Mantysaari:2023prg,Kryshen:2023bxy}. In particular, this explains the destructive interference at $t=0$ in ultraperipheral A+A collisions, as first proposed by  Klein and Nystrand~\cite{Klein:1999gv} and verified by the STAR Collaboration shortly thereafter~\cite{STAR:2002caw}, and recently in $J/\psi$ photoproduction~\cite{STAR:2023nos,STAR:2023nos}.
While the models capture key features of the data (including the approximate impact parameter dependence measured by ALICE), a fully quantitative description is lacking; in addition to theoretical uncertainties in modeling nonperturbative features of the process, a clean separation of coherent and incoherent contributions is challenging. 

In this paper, we will develop a formalism that naturally incorporates the dynamics included in the aforementioned models but demonstrates further that a novel variant of the Cotler-Wilczek $E^2 I^2$ mechanism can further help understand the full systematics of the exclusive vector meson data. The essential element in this proposed interferometric analysis of data is the phase information contained in the entanglement of the vector meson wavefunctions with their daughter particles. Thus while the quantum interference effects from the entanglement of nuclear wavefunctions at very low $|\v q_\perp|$ rapidly vanish with increasing $|\v q_\perp|$, the role played by the vector mesons as an ``entangling filter" in resonant decays persists at the larger $|\v q_\perp|$ corresponding to strong interaction scales. We emphasize that this $E^2 I^2$ mechanism is not unique to UPCs. It also provides an interferometric tool, for all such exclusive resonant decays,  that are applicable to deeply inelastic scattering (DIS) experiments in electron-nucleus collisions at the future Electron-Ion Collider (EIC)~\cite{Accardi:2012qut,Aschenauer:2017jsk,AbdulKhalek:2021gbh}. 

A noteworthy aspect of our formalism  suggested by the results of our study is that $E^2 I^2$ allows one to extract fundamental model independent information on the strong interactions. In particular, we show our results on exclusive $\rho$-meson photoproduction are sensitive to the distribution of $C=1$ color singlet ``pomeron" configurations of gluons in nuclei that dominate total cross-sections in QCD at high energies. This opens the door to determining whether the pomeron flux in a nucleus is universal, as befitting an object believed to carry the quantum numbers of the vacuum, or rather, is ``shadowed" by the presence of other gluon configurations in a nucleus in a  manner similar to that of gluon distributions in nuclei. Our results also allow one to explore the sensitivity of the UPC data to the spin structure of pomeron couplings to hadrons, which is a matter of some debate. Another  intriguing possibility is that $E^2 I^2$ can provide a further tool (from exclusive measurements of $C=+1$ heavy quarkonium states) in searches for the pomeron's $C=-1$ partner, the odderon.

A further novel aspect of our work, consistent with its $E^2 I^2$ underpinning, is our treatment of the exclusive decay of the $\rho$-meson as an entangled state of $\pi^+\pi^-$. This approach is different from the usual 
spin density matrix formalism~\cite{Bauer:1977iq,Schilling:1969um} often employed to describe exclusive decays. In our view, the $E^2 I^2$ language of entangled spin and angular momentum decomposed Fock states, and the corresponding production and decay amplitudes, is better suited to extract differential information (from increasingly precise collider data) on a number of subtle and little understood features of the strong interactions. An example where this framework can be applied is to explore the validity~\cite{Cisek:2022yjj} of the phenomenologically successful s-channel helicity conservation (SCHC) formalism~\cite{Gilman:1970vi} for exclusive vector meson decays. For an excellent review that incorporates a detailed study of SCHC (and its violation) in HERA data, we refer the reader to \cite{Ivanov:2004ax}. 

The paper is organized as follows. We begin  Section~\ref{section:2} by outlining the framework that will be employed in our computation of the exclusive photoproduction of $\rho$-mesons in UPCs that subsequently decay to $\pi^+\pi^-$ pairs. The general setup of the overall computation is briefly outlined in section~\ref{section:2A}. The explicit computation of the $\rho$-meson amplitude is performed in section~\ref{section:2B} with the final result obtained in Eq.~\eqref{eq:final-12-rho-amplitude}. In section~\ref{section:2C}, we discuss the computation of amplitude for $\rho\rightarrow \pi^+\pi^-$, demonstrating the key role in the decay of the relative phase of the helicity states of the $\rho$-meson. We put together our results in section~\ref{section:3}, where we compute the average number of correlated pairs produced in the collision. Specifically,  analytical results for the coherent exclusive cross-section are provided and analyzed in section~\ref{section:3A}. In section~\ref{subsection:3B}, we extend the analysis to the total exclusive cross-section that includes both coherent and incoherent contributions. We discuss qualitative features of the results and how they compare to those seen in the experimental data. The application of the formalism to other exclusive final states is discussed briefly in section~\ref{subsection:3C}. In section~\ref{subsection:3D}, we discuss our framework and results in the context of recent UPC literature on exclusive final states.  We conclude in section~\ref{section:4} with a summary of our results and an outlook towards future measurements both in UPCs and at the EIC. We will also discuss briefly the possibility that the techniques discussed here can be applied to extract specific interferometric information in cold atom experiments. The paper contains two appendices. In Appendix~\ref{Appendix:A}, we present for completeness, a brief synopsis of the essentials of HBT interferometry. Likewise, the Cotler-Wilczek idea is briefly summarized in Appendix~\ref{Appendix:B}. 

\section{Exclusive production of $\pi^\pm$-pairs in ultrarelativistic nuclear collisions
}
\label{section:2}

In this section, we compute the exclusive production of $\pi^\pm$-pairs from the decay of $\rho$-mesons produced in ultraperipheral ultrarelativistic nuclear collisions (UPCs). Since this is an exclusive process, $\rho$-mesons are produced in the reaction 
\begin{equation}
\gamma (A_1) + {\mathbb P} (A_2) \rightarrow \rho (A_2)\rightarrow \pi^+ \pi^-\,.
\end{equation}
Here $\gamma$ denotes an equivalent quasi-real Weizs\"{a}cker-Williams photon from the 
boosted electromagnetic field of nucleus $A_1$ which scatters off a color singlet configuration in nucleus $A_2$ (denoted here by $\mathbb P$) to produce the 
$\rho$ meson. This color singlet configuration can be understood as the $t$-channel exchange of an object called the pomeron carrying vacuum quantum numbers (zero electric charge, flavor, and baryon number). A phenomenological description of elastic, diffractive, and exclusive scattering in terms of pomeron exchange has long provided a successful description of soft processes in strong interactions~\cite{Forshaw:1997dc}. 

\begin{figure}
\centering\includegraphics[width=0.9\linewidth]{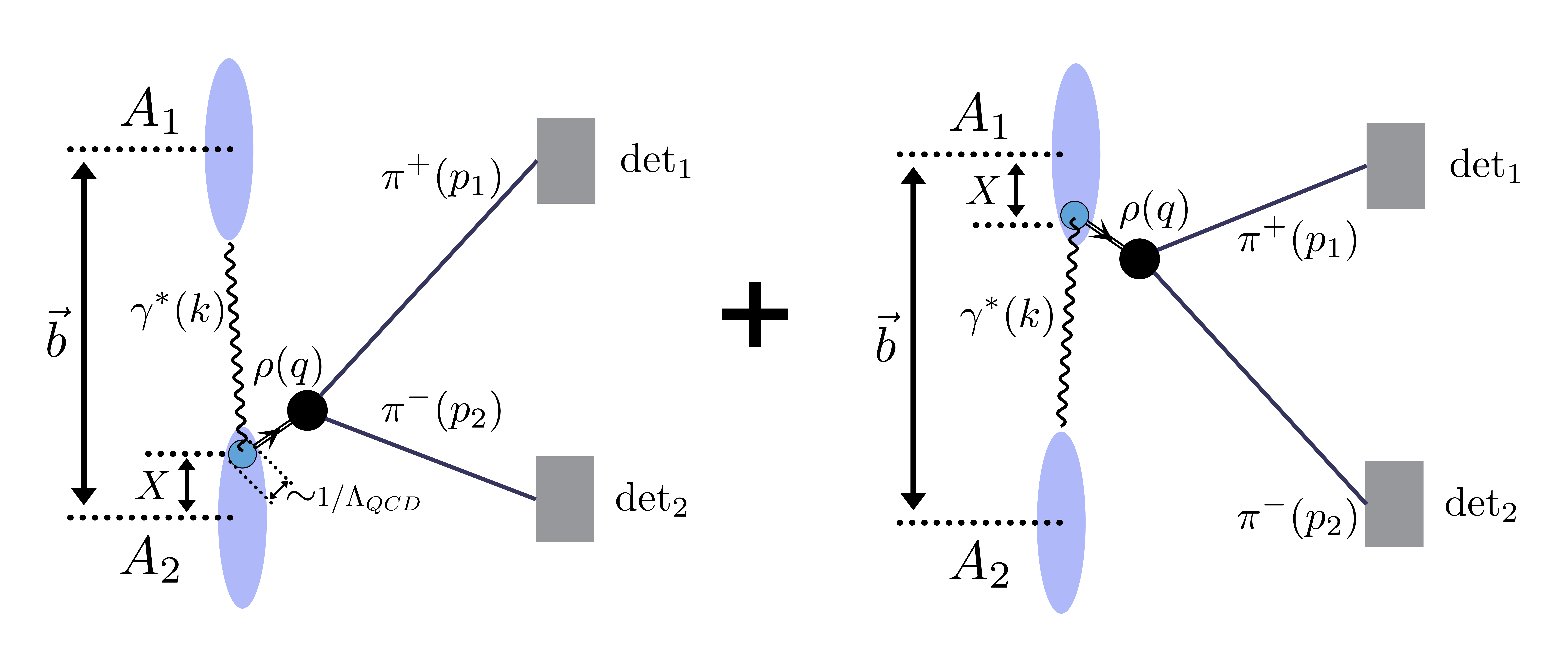}
\caption{Resonant production of $\pi^\pm$-pairs from  exclusive $\rho$-meson decay in ultraperipheral ultrarelativistic nuclear collisions at a given impact parameter $\v b$. The left figure depicts the production of a $\rho$-meson within nucleus 2 through the fusion of a Weizs\"{a}cker-Williams photon from nucleus 1 that scatters off a color singlet (pomeron) configuration localized within a patch of size $1/\Lambda_{\rm QCD}$ at a distance specified by $\v X$ from the center of nucleus 2. The right figure depicts the reverse process.
}
 \label{fig:UPC}
\end{figure}

Since the $\rho$ meson generated in a strong interaction process has a limited interaction range, its production is localized to the diameter of nucleus $A_2$. 
Note that an identical process occurs with $A_1\leftrightarrow A_2$. They are both illustrated in  Fig.~\ref{fig:UPC}. The $\rho$-mesons produced in this process decay rapidly on strong interaction time scales\footnote{As emphasized in \cite{Klein:2002gc}, the decay width of the $\rho$-meson corresponding to strong interaction time scales does not mean that its wavefunction has collapsed on these time scales. The information on the decay probability into the $\pi^+\pi^-$ pair, and of their relative $P$-wave phase, are both extracted in the measurement of the $\pi^\pm$-pairs.}. In contrast, the $\pi^\pm$-pairs are stable on strong interaction time scales and their wavefunctions become delocalized on the much longer time scales of the measurement. Since the source for the pairs is not distinguishable, the pair production cross-section results from a linear combination of their emission amplitudes from the two nuclei. 

In QCD, the simplest description of the pomeron is a color singlet combination of two gluons~\cite{Low:1975sv,Nussinov:1975mw}; however this description is understood to be oversimplified. The extrapolation of the pomeron's $t$-channel Regge trajectory to positive momentum transfers does not reveal any clearly identifiable states though there have been glueball candidates whose masses lie close to this trajectory~\cite{Meyer:2004jc}. Indeed, even assuming the pomeron to be a quasiparticle-like object, its spin has not been unambiguously settled between whether it is a spin-1 like the photon or a spin-2 exchange like that of a tensor glueball~\cite{Kopeliovich:1989hp,Goloskokov:1995uh,Nachtmann:1997hei,Donnachie:1984xq,Hagiwara:2020mqb}. 

Of further relevance to our discussion is the question of whether the pomeron is a coherent excitation of the nucleus on distance scales $R$, where $R$ is the nuclear radius, or localized on distance scales $1/\Lambda_{\rm QCD} << R$, with  $\Lambda_{\rm QCD}$ denoting the fundamental momentum scale of QCD. Evidence for the former can be inferred from the observed width of the diffractive peak in scattering off nuclei and the shrinkage of this peak with increasing energy~\cite{Kowalski:2008sa}. However one can also interpret this as a combination of scattering off a black disc core (where the $S$-matrix goes to zero) and an annular corona region of width $1/\Lambda_{\rm QCD}$~\cite{Kovner:2002xa,Nussinov:2008nz,Gotsman:2014pwa,Werner:2023zvo}. 

A weak coupling (albeit nonperturbative) picture consistent with this ``core+corona" description can be realized  for large nuclei at high energies is that of the Color Glass Condensate (CGC) effective field theory (EFT). In this EFT, color charge distributions are screened on distance scales $1/Q_S(b)\ll 1/\Lambda_{\rm QCD}$, where $Q_S(b)$ is the saturation scale for gluons at a given impact parameter $b$~\cite{Iancu:2003xm,Gelis:2010nm,Kovchegov:2012mbw}. Coherent diffractive/exclusive scattering in this picture corresponds to {\it averaging amplitudes} over color charge distributions in patches of maximal size $1/\Lambda_{\rm QCD}$, with weights that ensure color singlet configurations are dominant. The modulus squared of amplitudes thus constructed gives the coherent diffractive/exclusive cross-section. In contrast, for the fully inclusive exclusive vector meson cross-section, one first takes the squared modulus of amplitudes computed for the given localized color charge configurations and then finally performs the weighted average over the cross-sections~\cite{Kovchegov:1999kx}. The incoherent exclusive cross-section is the difference between the two cross-sections. 

Since the dynamics of interest to us in UPCs occurs at soft momentum scales $\ll \Lambda_{\rm QCD}$, and is therefore intrinsically nonperturbative, we will attempt to keep our computation as general as possible without relying unduly on a specific model.
The only assumption we will make is that the $\gamma+{\mathbb P}\rightarrow \rho$ cross-section occurs in patches of size $\sim 1/\Lambda_{\rm QCD}$ at positions within the nucleus (denoted by $\v{X}$ in Fig.~\ref{fig:UPC}), in keeping with our general understanding of the confinement of parton degrees of freedom.

In the intermediate steps of our computation however, we will rely on the CGC formulation because it (uniquely among various approaches in the literature on exclusive final states) provides a spacetime picture that allows us to keep track of the relevant phase information. However, we will rephrase our results so that they are compatible with other approaches in the literature.  We will see that these minimal assumptions are sufficient to obtain a model independent $E^2 I^2$ interpretation of the UPC data on exclusive $\pi^\pm$-pair production. Therefore more quantitative ``precision" comparisons with data, and similar results for other vector mesons (such as exclusive 
$\phi\rightarrow K^+ K^-$, $J/\psi\rightarrow e^+ e^-$,$\cdots$) can,  via $E^2 I^2$, provide deeper insight into the fundamental nature of pomeron dynamics. A further possibility we alluded to previously is that this novel interferometric tool can provide independent evidence for the existence of the pomeron's C-odd odderon partner (a simple model being the color singlet exchange of three gluons~\cite{Lukaszuk:1973nt}) to confirm recent claims regarding the observation of this exchange at the LHC~\cite{TOTEM:2017sdy}.

\subsection{Setting up the problem}
\label{section:2A}
As illustrated in Fig.~\ref{fig:UPC}, the initial state for our computation of $\pi^\pm$-pair production is characterized by  $|A_1,A_2,\v b \rangle $, where $\v b$ is the impact parameter of the scattering corresponding to the distance between the centers of the two nuclei. The amplitude to produce a $\rho$-meson with transverse momentum $\v q$ is obtained by computing the overlap of the $\rho$-meson state with a photon+pomeron state $|\gamma(\v b), {\mathbb P}(\v X) \rangle $ that is projected out from the nuclear wavefunctions. To obtain the $\pi^\pm$-pair amplitude, we have to subsequently compute the overlap of $\langle\rho(\v q)|\pi^+ \pi^-,\v p_1,\v p_2\rangle$, with $\v q = \v p_1+\v p_2$.
For the amplitude specified in Fig.~\ref{fig:UPC}, $\v b$ denotes the position of the photon relative to the center of nucleus $A_1$ and likewise, $\v X$ denotes the position of the pomeron ``patch" within nucleus $A_2$. 
The exclusive pair production amplitude can then be expressed as 
\begin{equation}
\label{eq:master-eq}
\begin{split}
M_{A_1 A_2\rightarrow \pi^+\pi^-}(\v p_1,\v p_2,\v b)= M_{A_1 A_2\rightarrow \rho}(\v q,\v b)\,M_{\rho \rightarrow \pi^+\pi^-}(\v q,\v p_1,\v p_2)\,,
\end{split}
\end{equation}
with 
\begin{equation}
\label{eq:gamma-pomeron}
\begin{split}
    M_{A_1A_2\rightarrow\rho}(\v q,\v b)=\int_{\v X}\langle A_1,A_2, \rho(\v q)|(U_{\rm dipole}-1)|\gamma(\v k), {\mathbb P}(\v X)\rangle\,\langle \gamma(\v k), {\mathbb P}(\v X)| (U_{\rm photon}-1)(U_{\rm Pom.}-1)|A_1,A_2,\v b  \rangle \,.
    \end{split}
\end{equation}

Here $M_{A_1A_2\rightarrow\rho}$ is the exclusive amplitude for producing a $\rho$-meson at a location $\v b$ relative to the center of nucleus $A_1$.  Further,  $\mathcal{M}_\rho(\v q,\v p_1,\v p_2)$ is the amplitude for the $\rho$-meson with transverse momentum $\v q$ to decay into the $\pi^+\pi^-$-pair, denoted as
\begin{equation}
    \mathcal{M}_{\rho \rightarrow \pi^+\pi^-}(\v q,\v p_1,\v p_2)= \langle A_1,A_2,\rho(\v q)|(U_{\rm frag.}-1)|\pi^+ \pi^-,\v p_1,\v p_2 \rangle \,.
\end{equation}

The terms $U_{\rm photon}$, $U_{\rm Pom.}$, $U_{\rm dipole}$ and $U_{\rm frag.}$ are respectively the unitary operators corresponding to terms in the QED+QCD Hamiltonian generating the photon, the pomeron, the $\rho$-meson and the $\pi^+\pi^-$-pair through their action on the relevant Fock states~\cite{Bjorken:1970ah}. They generate the flux factors of the photon and the pomeron, the matrix element for the $\gamma {\mathbb P}\rightarrow \rho$ process, and that for the fragmentation of $\rho\rightarrow \pi^+\pi^-$. 

Apart from the photon flux factor, these are intrinsically nonperturbative quantities in the kinematic regime of interest. It would therefore be valuable to express the results in as model independent a manner as possible. Since our interest is primarily in the phase information and the spatial distribution of the pomeron in the nucleus, it will be important to treat the appropriate phases and polarization structures with care.  In what follows, we will for compactness replace $ M_{A_1 A_2\rightarrow\rho}\rightarrow M_{{12}\rightarrow\rho}$
(and $M_{A_2 A_1\rightarrow\rho}\rightarrow M_{{21}\rightarrow\rho}$).

\subsection{The $\gamma\,{\mathbb P}\rightarrow \rho$ amplitude}
\label{section:2B}
We will employ prior computations \cite{Gelis:2001dh,Dumitru:2018vpr} in the CGC EFT framework  to describe the exclusive production of a $q{\bar q}$ pair in ultraperipheral nuclear collisions. We are of course interested here in the exclusive production of a $\rho$-meson; in this approach, it follows from the nonperturbative projection of the $q{\bar q}$-dipole on the $\rho$-meson state\footnote{An alternative nonperturbative description is through the phenomenologically motivated Vector Meson Dominance (VMD) hypothesis~\cite{Gounaris:1968mw}. For our purposes, the specifics of the overlap of the photon and $\rho$-meson wavefunction will not matter at this stage, and can be introduced later.}. As noted previously, the same formalism can be applied to the exclusive production of heavy quarkonia, for which NRQCD is applicable in describing the overlap of the dipole and Onium states, in which case, a CGC+NRQCD framework can be applied~\cite{Kang:2013hta}. 

The general form for the amplitude in Eq.~\ref{eq:gamma-pomeron} for the exclusive production of a $\rho$-meson can be reexpressed as 
\begin{eqnarray}
\label{eq:full-amplitude}
        M_{12\rightarrow \rho}^{\lambda_\gamma \lambda_\rho}(\v q,\v b) &=& \int \frac{d^4\v l}{(2\pi)^4}\frac{d^4\v p}{(2\pi)^4}\, \delta^{(4)}(\v l + \v p - \v q)\, {\cal N}_{q{\bar q}\rightarrow \rho}^{\lambda_1\lambda_2 \lambda_\rho}(\v l, \v p; \v q)
        \int \frac{d^2\v k_\perp}{(2\pi)^2}\frac{d^2\v K_\perp}{(2\pi)^2}\,\frac{d^2{\v \Delta K}_\perp}{(2\pi)^2}\,\nonumber\\
         &\times& \delta^{(2)}(\Delta \v K_\perp + \v k_\perp - \v p_\perp - \v l_\perp)\,M_{\gamma A_2\rightarrow q{\bar q}}^{\lambda_\gamma \lambda_{\mathbb P}\lambda_1\lambda_2}(\v l, \v p| \v k_\perp, \v K_\perp,{\v \Delta K}_\perp ; \v b)  \, .
\end{eqnarray}
Here ${\cal N}_{q{\bar q}\rightarrow \rho}^{\lambda_1\lambda_2 \lambda_\rho}(\v l, \v p; \v q)$ represents the nonperturbative overlap of the $q{\bar q}$ and $\rho$-meson wavefunctions, with $\lambda_{1}$ ($\lambda_2$) denoting the helicity of the quark (antiquark), which will be summed over. Further, $\lambda_\gamma$, $\lambda_{\mathbb P}$, and $\lambda_\rho$, respectively, represent the helicities of the photon, the pomeron, and the $\rho$-meson. 

As noted, the amplitude $M_{\gamma A_2\rightarrow q{\bar q}}^{\lambda_\gamma \lambda_{\mathbb P}\lambda_1\lambda_2}(\v l, \v p| \v k, \v k_1 \v k_2; \v b)$ was explicitly computed in the CGC EFT. Suitably adapted to our problem, it can be expressed as~\cite{Gelis:2001da,Gelis:2001dh,Dumitru:2018vpr}  
\begin{eqnarray}
\label{eq:color-averaged-amplitude}
M_{\gamma A_2\rightarrow q{\bar q}}^{\lambda_\gamma \lambda_{\mathbb P}\lambda_1\lambda_2}(\v l, \v p| \v k_\perp, \v K_\perp, {\v \Delta K}_\perp;\v b)
\equiv {\tilde \rho}_{\rm EM}(\v k_\perp;\v b)\, \langle M_{\gamma A_2\rightarrow q{\bar q}}^{\lambda_\gamma \lambda_{\mathbb P}\lambda_1\lambda_2}(\v l, \v p| \v k_\perp, \v K_\perp,\v {\Delta K}_\perp)[\rho_{QCD}]\rangle_{\rho_{\rm QCD}}\,,
\end{eqnarray}
where 
\begin{eqnarray}
\label{eq:color-scattering-amplitude}
    \langle M_{\gamma A_2\rightarrow q{\bar q}}^{\lambda_\gamma \lambda_{\mathbb P}\lambda_1\lambda_2}(\v l, \v p| \v k_\perp, \v K_\perp, \v {\Delta K}_\perp)[\rho_{QCD}]\rangle_{\rho_{\rm QCD}} &=& \int d^2 \v X \,d^2 \v r\,\, e^{i \frac{\v r}{2} \cdot \v {K_\perp}}\,
    e^{i\v X \cdot \v {\Delta K_\perp} }  \left[1 - \frac{1}{N_c}{\rm Tr} \langle U_\rho (\v X+ \frac{\v r}{2})U_\rho^\dagger(\v X - \frac{\v r}{2})\rangle_{\rho_{\rm QCD}}\right]\nonumber \\
    &\times& {\bar M}_{\gamma A_2\rightarrow q{\bar q}}^{\lambda_\gamma \lambda_{\mathbb P}\lambda_1\lambda_2}(\v k_\perp,\v K_\perp, \v {\Delta K}_\perp; \v p, \v l) \,.
\end{eqnarray}

Here $\v q$ denotes the $\rho$-meson four-momentum, and $\v k$, $\v p$ and $\v l$ represent those of the photon, the quark and the antiquark, respectively. Further, $\v k_1$ and $\v k_2$ denote the momenta of gluons in the nuclear wavefunction that scatter off the photon to produce the $\rho$-meson. As we will discuss later, we will focus our discussion on UPC measurements at central rapidities ($Y=0$) that simplifies the problem considerably. While in principle the photons and gluons contributing to $\rho$-meson production in this kinematics can have large equal but opposite longitudinal momentum in the center-of-mass frame, most of the glue in the nuclear wavefunction have very small longitudinal momenta (small momentum fractions $x$). Thus the kinematics relevant to central rapidity UPC measurements corresponds to $k_z\sim k_{1z}\sim k_{2z}\sim x\,P_z\leq 1$ GeV. 

Since we are interested in an exclusive process, we note that the gluons must be in a color singlet configuration with net momentum $\Delta\v K_\perp=\v k_1 +\v k_2$, and relative  momenta $\v { K}_\perp = \v k_1 - \v k_2$. Because $\v {\Delta K}_\perp$ is conjugate to the impact parameter $\v X$ of the color singlet patch in the nucleus, it corresponds to a momentum $\sim 1/2R$, where $R$ is the radius of the nucleus. On the other hand, $\v K_\perp$ is conjugate to the size of the 
$q\bar q$-dipole/$\rho$-meson. Unlike deeply inelastic scattering (DIS) where the size of the dipole is set by the photon virtual momentum, the size of the dipole in UPCs is set by the inverse of the momentum transfer from the target to the dipole. While for $\rho$-meson production, the amplitude will be dominated by $\v K_\perp\sim \Lambda_{\rm QCD}$, there will be (power suppressed) contributions from larger values of $\v K_\perp$ corresponding to smaller dipoles.

Before we proceed to discuss Eq.~\eqref{eq:color-scattering-amplitude} further, we note that the electromagnetic contribution to the amplitude in Eq.~\eqref{eq:color-averaged-amplitude} can be represented as\footnote{A subtle point is that $F_{\rm QED}$ is in light cone gauge, and this must be kept in mind in computing the amplitude in \cite{Gelis:2001da,Gelis:2001dh}. It is related to the covariant gauge expression by a simple gauge transformation~\cite{Jackiw:1991ck,Baltz:1998zb}. The electromagnetic form factor appearing in the cross-section is of course gauge invariant.}
\begin{equation}
{\tilde \rho}_{\rm EM}(\v k_\perp;\v b)= 
\int d^2\v {\bar b} 
\left[e^{-ie \Lambda(\v b-\v {\bar b})}-1\right]\, 
e^{i\v k_\perp\cdot \v {\bar b}}
\equiv 
e^{i\v k_\perp\cdot \v b}\,
F_{\rm QED}(\v k_\perp)\,.
\end{equation}
Here $\Lambda(x_\perp) = \int d^2 y_\perp G_0(x_\perp - y_\perp)\, \rho_{\rm EM}(y_\perp)$, where $\rho_{\rm EM}(x_\perp)$ is the electromagnetic charge distribution in the nucleus and $G_0(x_\perp-y_\perp)= \frac{1}{2\pi} \ln(|\v x_\perp - \v y_\perp|)$ is the two-dimensional Coulomb propagator. For a ``pointlike" coherent nuclear source, ${\tilde \rho}_{\rm EM}(\v k;\v b)= \frac{\v b}{|\v b|}e^{i\v k_\perp \cdot \v b}\, \frac{4\pi Z\alpha}{\v k_\perp^2}$, where $Z$ is the nuclear charge and $\alpha$ is the electromagnetic fine structure constant\footnote{Note that in our convention the charge $e$ in the matrix element is absorbed in the form factor.}. 

This momentum distribution is of course the Weizs\"{a}cker-Williams (WW) equivalent photon distribution. In the cross-section, we will replace it by the electromagnetic form factor of a boosted nucleus when $\gamma\rightarrow \infty$. Most relevant for our discussion however is the fact that the WW photons are linearly polarized in the direction of the impact parameter, representing the direction of the electric field of the nucleus~\cite{Li:2019yzy,Klein:2020jom}, with unit  polarization vector  ${\v \varepsilon}^{\lambda_\gamma}=\frac{\v b}{|\v b|}$.  (They are transversely polarized relative to the direction of the beam.) In computing the sum of the amplitudes $M_{12\rightarrow \rho} + M_{21\rightarrow \rho}$, we will see that this sum will cancel because the polarizations are equal and opposite when the $\rho$-meson momentum $\v q\rightarrow 0$. 

We now return to the computation of  Eq.~\eqref{eq:color-scattering-amplitude}. 
In the CGC EFT, the gluons in the ultrarelativistic nuclei are sourced locally by classical color charges\footnote{The gluons couple coherently to a large number of color charges that typically live in a high dimensional color representation.} and their interactions are described by the lightlike Wilson lines\footnote{The color sources are path ordered in rapidity, which is important when one computes color averages.} 
\begin{equation}
\label{eq:color-rotation}
    U(\v x_\perp) = P\exp\left( i g \, \frac{\rho(\v x_\perp)}{\nabla_\perp^2}\right)\,,
\end{equation}
where $\rho(\v x_\perp)$ represents\footnote{Both the color charge density and the $\rho$-meson are represented by the Greek symbol but we anticipate that this will not cause confusion due to their very different useage.} the color charge per unit area localized at a position $\v x_\perp$ in the transverse plane of the nucleus. Gluon dynamics in a high energy process occurs on much shorter time scales relative to the lifetimes of the color sources, which enables one to treat them as static stochastic sources~\cite{McLerran:1993ni,McLerran:1993ka}. For exclusive {\it coherent} final states, one averages over color singlet patches within a nucleon, and then over nucleon distributions in the nuclei, at the level of the amplitude, as denoted by $\langle \cdots \rangle_{\rho_{\rm QCD}}$ in Eq.~\eqref{eq:color-averaged-amplitude}. This ensures that the gluons within a color singlet ``patch" $\bf X$ ($\sim 1/\Lambda_{\rm QCD}$) within the nucleus in the amplitude are emitted in a color singlet configuration with momentum $\v K_\perp$, with the nucleus remaining intact. It is this color averaged configuration we will understand henceforth to be the pomeron. Thus  ${\bar M}(\v k,\v K_\perp, \v {\Delta K}_\perp; \v p, \v l)$ in Eq.~\eqref{eq:color-scattering-amplitude} represents the coherent photon+pomeron$\rightarrow q{\bar q}$ scattering amplitude. 

Generalizing the results of \cite{Dumitru:2018vpr}, Eq.~\eqref{eq:color-scattering-amplitude} can be rewritten as 
\begin{eqnarray}
\label{eq:pomeron-flux}
\langle M_{\gamma A_2\rightarrow q{\bar q}}^{\lambda_\gamma \lambda_{\mathbb P}\lambda_1\lambda_2}(\v l, \v p| \v k_\perp,\v K_\perp, {\v \Delta K}_\perp)\left[\rho_{\rm QCD}\right]\rangle_{\rho_{\rm QCD}} = \tilde P({\v \Delta K}_\perp, K_\perp)\,{\bar M}_{\gamma {\mathbb P} \rightarrow q {\bar q}}^{\lambda_\gamma \lambda_{\mathbb P}\lambda_1\lambda_2}(\v k_\perp,\v K_\perp,  {\v \Delta K}_\perp; \v p, \v l)\,,
\end{eqnarray}
where we have integrated over the relative dipole size $\v r$ and the pomeron impact parameter $\v X$. The expression ${\tilde P}$ is simply proportional\footnote{It has the form $\frac{1}{K_\perp^4} G(-\frac{\v K_\perp - {\v \Delta K}_\perp}{2},\frac{\v K_\perp- {\v \Delta K}_\perp}{2})$, for $\v K_\perp \gg {\v \Delta K}_\perp$, where $G$ is the  pomeron form factor. Despite the strong inverse power law dependence on $\v K_\perp$, the integral over $\v K_\perp$ is infrared safe since the pomeron form factor goes to zero with $\v K_\perp \rightarrow 0$ as a consequence of color neutrality. } to the general definition of the pomeron form factor given in \cite{Dumitru:2018vpr}.  Expressions for the amplitude  
${\bar M}(\v k,\v K_\perp, \v {\Delta K}_\perp; \v p, \v l)$ and its 
squared modulus can be computed straightforwardly \cite{Gelis:2001da,Gelis:2001dh}. Substituting Eq.~\eqref{eq:pomeron-flux} in  Eq.~\eqref{eq:color-averaged-amplitude}, and then the latter into Eq.~\eqref{eq:full-amplitude}, we obtain
\begin{eqnarray}
\label{eq:finalA-12-rho-amplitude}
    &&
    M_{12\rightarrow \rho}^{\lambda_\gamma \lambda_{\mathbb P}\lambda_\rho}(\v q,\v b) = \int \frac{d^4\v l}{(2\pi)^4}\frac{d^4\v p}{(2\pi)^4}\, \delta^{(4)}(\v l + \v p - \v q)\, {\cal N}_{q{\bar q}\rightarrow \rho}^{\lambda_1\lambda_2\lambda_\rho}(\v l, \v p; \v q)\,
        \int \frac{d^2\v k_\perp}{(2\pi)^2}\frac{d^2\v K_\perp}{(2\pi)^2}\,\frac{d^2{\v \Delta K}_\perp}{(2\pi)^2}\,\nonumber\\
         &\times& \delta^{(2)}(\Delta\v K_\perp + \v k_\perp - \v p_\perp - \v l_\perp)\,
 e^{i\v k_\perp\cdot \v b}F_{\rm QED}(\v k_\perp)\,
{\tilde P}({\v \Delta K}_\perp, \v K_\perp)\,{\bar M}_{\gamma {\mathbb P} \rightarrow q {\bar q}}^{\lambda_\gamma \lambda_{\mathbb P}\lambda_1\lambda_2}(\v k,\v K_\perp, \v {\Delta K}_\perp; \v p, \v l)\,.
\end{eqnarray}
This can be further simplified to read
\begin{eqnarray}
\label{eq:finalB-12-rho-amplitude}
    M_{12\rightarrow \rho}^{\lambda_\gamma \lambda_{\mathbb P}\lambda_\rho}(\v q,\v b) &=& e^{i\v q\cdot \v b}\,
        \int \frac{d^2\v K_\perp}{(2\pi)^2}\,\frac{d^2{\v \Delta K}_\perp}{(2\pi)^2}\,
 \,\,e^{-i{\v \Delta K}_\perp\cdot \v b}\,F_{\rm QED}(\v q_\perp - \Delta\v K_\perp)\,
{\tilde P}({\v \Delta K}_\perp, K_\perp)\,\nonumber \\
&\times& \int \frac{d^4\v {\Delta q}}{(2\pi)^4}\,\,{\bar M}_{\gamma {\mathbb P} \rightarrow q {\bar q}}^{\lambda_\gamma \lambda_{\mathbb P}\lambda_1\lambda_2}(\v q_\perp -\Delta\v K_\perp,\v K_\perp, \v {\Delta K}_\perp; \v q ,\v \Delta q)\,\,{\cal N}_{q{\bar q}\rightarrow \rho}^{\lambda_1\lambda_2\lambda_\rho}({\v \Delta q}; \v q)\,.
\end{eqnarray}
Thus, we see that the amplitude for the exclusive production of a $\rho$-meson can be expressed as a convolution of the photon and pomeron flux factors times the amplitude for $\gamma {\mathbb P}\rightarrow \rho$ represented by the second line of the expression above. Here $\v {\Delta q}=\v l- \v p\sim \Lambda_{\rm QCD}$. 

An illustrative way to write the above expression is via the partial Fourier transform of the pomeron form factor: 
\begin{eqnarray}
\label{eq:pomeron-FT}
    {\tilde P}(\v {\Delta K}_\perp, \v K_\perp)= \int_{|\v X|<R} d^2 \v X \, e^{i\v X\cdot {\v {\Delta K}_\perp}}\, P(\v X, \v K_\perp) \,. 
\end{eqnarray}
The function on the r.h.s therefore represents the pomeron form factor in a color singlet patch $\v X$ within a nucleus. In the forward limit $\v {\Delta K}_\perp=0$, it is the average over the pomeron distribution over all the color singlet patches in the nucleus. 
If the size of the patch is that of a nucleon, this would be the form factor of the nucleon; one can however choose to be agnostic and treat it as a scale to be extracted from data. As we will see, this possibility is a consequence of $E^2 I^2$. 

Substituting Eq.~\eqref{eq:pomeron-FT} in Eq.~\eqref{eq:finalB-12-rho-amplitude}, and rearranging terms, we obtain for the coherent exclusive scattering amplitude:
\begin{eqnarray}
\label{eq:finalB-12-rho-amplitude}
    M_{12\rightarrow \rho}^{\lambda_\gamma \lambda_{\mathbb P}\lambda_\rho}(\v q,\v b) &=&  e^{i\v q\cdot \v b}
       \int \frac{d^2\v K_\perp}{(2\pi)^2} \, \int_{|\v X|<R}\,d^2 \v X\,P(\v X, \v K_\perp)\,\int \frac{d^4\v {\Delta q}}{(2\pi)^4}\,{\cal N}_{q{\bar q}\rightarrow \rho}^{\lambda_1\lambda_2\lambda_\rho}({\v \Delta q}; \v q)\,
\,\nonumber \\
&\times& \int\frac{d^2\Delta \v K_\perp}{(2\pi)^2}\,e^{i{\v \Delta K}_\perp\cdot (\v X-\v b)}
F_{\rm QED}(\v q_\perp - \Delta\v K_\perp)\,
{\bar M}_{\gamma {\mathbb P} \rightarrow q {\bar q}}^{\lambda_\gamma \lambda_{\mathbb P}\lambda_1\lambda_2}(\v q_\perp -\Delta\v K_\perp,\v K_\perp; \v q ,\v \Delta q)\,\,\,.
\end{eqnarray}
Since $|\v X-\v b|$ is a very large scale corresponding approximately to the distance between two nuclei, the phase in the second line of this expression will oscillate rapidly setting the integral to zero except in a small region with support $|\v {\Delta K}_\perp| \sim 1/|\v R-\v b|$. This expression can be rewritten as,
\begin{eqnarray}
\label{eq:final-12-rho-amplitude}
    M_{12\rightarrow \rho}^{T \lambda_{\mathbb P}\lambda_\rho}(\v q,\v b) &=& \frac{\v b}{|\v b|}\, e^{i\v q\cdot \v b}\,F_{\rm QED}\Big(\v q_\perp - \frac{1}{|\v b-\v R|}\Big)\,
       \int \int\frac{d^2\Delta \v K_\perp}{(2\pi)^2}\,\frac{d^2\v K_\perp}{(2\pi)^2} \, \int_{|\v X|<R}\,d^2 \v X\,e^{i{\v \Delta K}_\perp\cdot \v X}\,P(\v X, \v K_\perp)\nonumber \\
&\times& \int \frac{d^4\v {\Delta q}}{(2\pi)^4}\,
{\bar M}_{\gamma {\mathbb P} \rightarrow q {\bar q}}^{T \lambda_{\mathbb P}\lambda_1\lambda_2}\Big(\v q_\perp -\v \Delta K_\perp,\v K_\perp; \v q ,\v \Delta q\Big)\,{\cal N}_{q{\bar q}\rightarrow \rho}^{\lambda_1\lambda_2\lambda_\rho}({\v \Delta q}; \v q).
\end{eqnarray}
Eq.~\eqref{eq:final-12-rho-amplitude} is our general result for the amplitude for a photon from nucleus 1 to produce a $\rho$-meson in nucleus 2. In writing this expression, we have taken into account the fact that the WW photons are transversely polarized ($\lambda_\gamma\rightarrow T$) with respect to the photon direction, with their polarization pointing along the impact parameter direction. 

There are several key features of this result which can {\color{blue} be} itemized as follows.
\begin{itemize}
\item 
Even though in the intermediate steps of the computation we employed the language and methods of the CGC EFT, the final expression is very general. This is attractive because it is doubtful that the CGC EFT is applicable in the context of exclusive $\rho$-meson production where the physics is likely intrinsically nonperturbative. The pomeron form factor $P(\v X, \v K_\perp)$ of a color singlet patch $\v X$ in the nucleus is employed in different frameworks; alternately, it can be extracted employing $E^2 I^2$ in UPCs (as we will discuss) and compared to extractions from other experiments. Further, the second line in Eq.~\eqref{eq:final-12-rho-amplitude} can be understood as the $\gamma {\mathbb P}\rightarrow \rho$ amplitude computed in differing nonperturbative frameworks. As noted earlier, a particularly intriguing possibility is that the spin structure of pomeron couplings in the amplitude can be extracted from $E^2 I^2$ measurements in UPCs~\cite{Ewerz:2013kda,Ewerz:2016onn,Britzger:2019lvc}. 
\item Since $|\v b-\v X|\sim |\v b|$,  we expect ${\bar M}_{\gamma {\mathbb P} \rightarrow q {\bar q}}^{T \lambda_{\mathbb P}\lambda_1\lambda_2}$ to have a very weak dependence on $\v X$. If the pomeron distribution in the nucleus is uniform ($P(\v X, \v K_\perp)\rightarrow P(\v K_\perp))$, then Eq.~\eqref{eq:final-12-rho-amplitude} is simply proportional to the area of the nucleus. The cross-section for fixed impact parameter is hence proportional to 
$A^{4/3}$ and the integral over all impact parameter has an $A^2$-dependence in the forward limit - which we expect for coherent vector meson production~\cite{Mantysaari:2017slo}. Of course a nontrivial dependence of $P(\v X, \v K_\perp)$ on $\v X$ would be very interesting providing novel insight into the pomeron flux in a nucleus, distinguishable from that of a nucleon. 
\item Our derivation is applicable to any coherent exclusive vector meson production process in UPCs. (With slight modifications, it is also applicable to DIS.) In particular, for heavy $C=-1$ Onia  ($J/\psi, \Upsilon,\cdots$) one can combine the perturbative expression for ${\bar M}_{\gamma {\mathbb P} \rightarrow q {\bar q}}$ and replace ${\cal N}_{q{\bar q}\rightarrow J/\psi, \Upsilon,\cdots}$ with long distance NRQCD matrix elements, as discussed in \cite{Kang:2013hta}. 
\item If the exclusive final state that has a resonant two particle decay is a $C=+1$ state, as is the case for $\eta_c$ and $\chi$-mesons, our formalism suggests that 
their amplitude is sensitive to the Odderon~\cite{Lukaszuk:1973nt} distribution $i{\cal O}(\v X, \v K_\perp)$ in a nucleus. Such exclusive measurements of the Odderon at collider energies have triggered much interest recently~\cite{Dumitru:2019qec,Benic:2023ybl,Benic:2024pqe}. From the perspective of $E^2 I^2$, $\chi_c$ measurements in the decay channel $J/\psi\, \gamma \rightarrow e^+ e^- \gamma$ would be the most promising, if challenging, measurements. 
\end{itemize}

Finally, one sees very simply from our result that 
\begin{equation}
\label{eq:final-21-rho-amplitude}
M_{21\rightarrow \rho}^{T \lambda_{\mathbb P}\lambda_\rho}(\v q,\v b) = -\,e^{-i2\,\v q \cdot \v b}\, M_{12\rightarrow \rho}^{T \lambda_{\mathbb P}\lambda_\rho}\,.
\end{equation}
This is because, as shown in Fig.~\ref{fig:UPC}, the photon originates from nucleus 2 and interacts with a pomeron in nucleus 1 to produce the $\rho$-meson. Since the photon is linearly polarized in the direction of the impact parameter, this amounts to just switching the sign of the polarization vector. Thus when the $\rho$-meson momentum $\v q\rightarrow 0$, $M_{21\rightarrow \rho}^{T \lambda_{\mathbb P}\lambda_\rho}(\v q,\v b)= - M_{12\rightarrow \rho}^{T \lambda_{\mathbb P}\lambda_\rho}(\v q,\v b)$, a result first obtained in \cite{Klein:1999gv} and clearly observed in experiment. 

\subsection{The $\rho\rightarrow \pi^+\pi^-$ amplitude}
\label{section:2C}
The final $|\pi^+\pi^-\rangle$ state that reaches the detector is from the resonant decay of the $\rho$-meson\footnote{Recall that the decay $\rho^0\rightarrow \pi^0+\pi^0$ if forbidden by charge-parity conservation in the strong interactions.}. Since the $\pi^+$ and $\pi^-$ are decay products of the same particle, their wavefunctions emerge from an entangled state~\cite{Klein:2002gc}. We will assume here a model of the  pomeron which satisfies s-channel helicity  conservation (SCHC) noted previously. In the SCHC framework, the $\rho$-meson inherits the linear polarization of the Weizs\"{a}cker-Williams photon\footnote{Note that DIS experiments indicate that the cross-section for the exclusive production 
of vector mesons goes to zero for longitudinally polarized photons as $Q^2\rightarrow 0$. In such measurements, the polarization directions of the $\rho$-meson are summed over and do not resolve by themselves what fraction of $\rho$-mesons are produced longitudinally or transversely polarized with respect to the beam direction.}. In other words, neglecting spin flip contributions, the $\rho$-meson is produced with its polarization in the impact parameter direction. Indications of the transverse polarization of $\rho$-mesons in diffractive photoproduction experiments first seen over 50 years ago~\cite{Ballam:1970qn} (most recently confirmed in precision experiments at GlueX~\cite{GlueX:2023fcq}) are consistent with the hypothesis of s-channel helicity conservation~\cite{Gilman:1970vi,Carlitz:1971ht}. Further analysis (for instance by triggering on the production of higher mass resonances~\cite{Klein:2016dtn} in UPCs) to examine this assumption more closely will be important to understand better the spin couplings of the pomeron and possible violations of the SCHC hypothesis~\cite{Cisek:2022yjj,Ivanov:2004ax}. 

\begin{figure}[H]
   \centering
\includegraphics[scale=0.4]{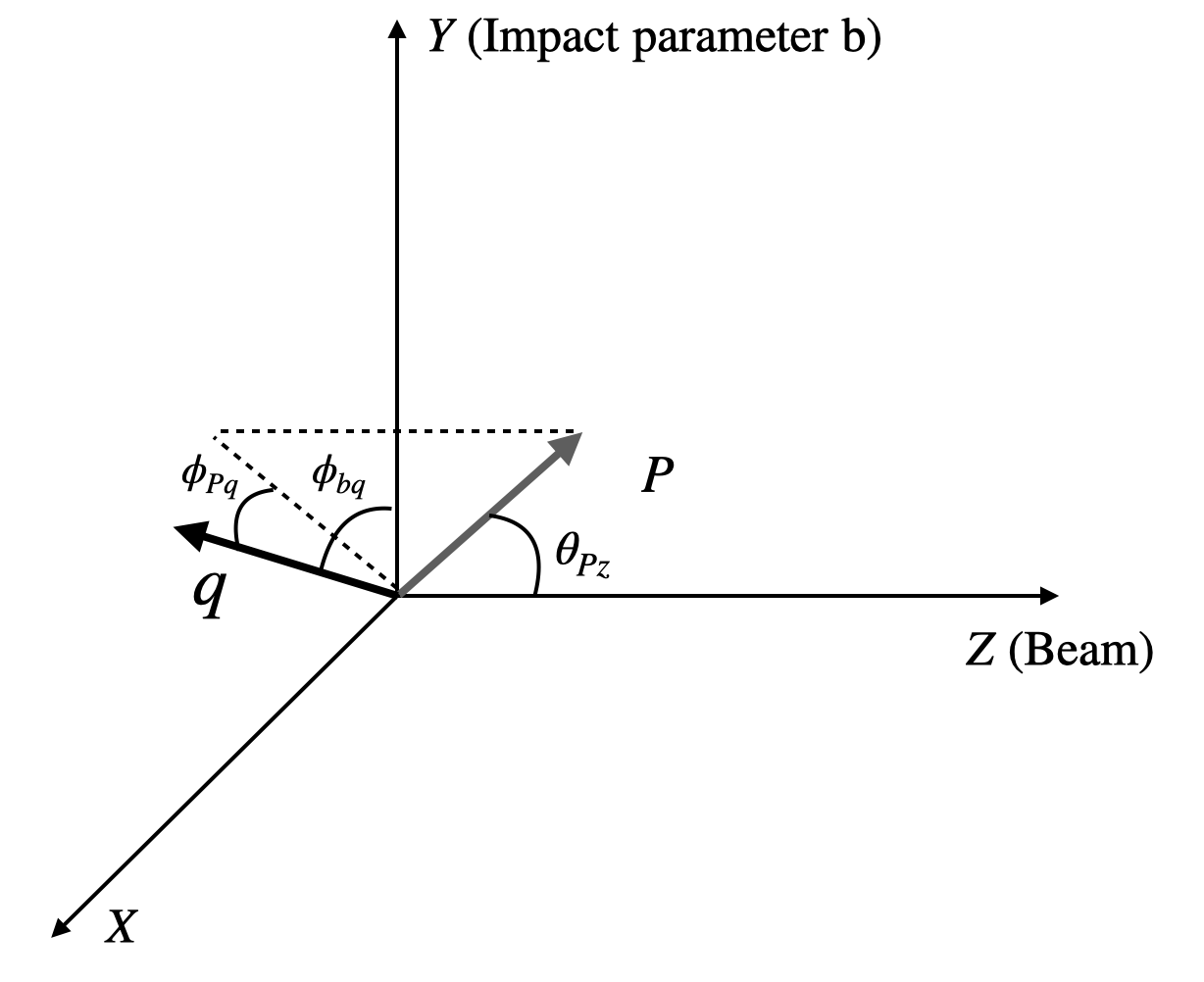}
\caption{The production and decay planes of the $\rho$-meson, reconstructed from the three-momenta $\v p_1$ and 
$\v p_2$ of the $\pi^\pm$ pair. Here $\v P = \v p_1 - \v p_2$ and $\v q= \v p_1 + \v p_2$. See text for further discussion.}
    \label{fig:angle}
\end{figure}

The decay amplitude for the $\rho\rightarrow \pi^+\pi^-$ can be expressed in terms of $\v q= \v p_1+ \v p_2$ and $\v P= \v p_1- \v p_2$, where $\v p_1$ and $\v p_2$ are respectively the $\pi^+$ and $\pi^-$ three-momenta.  Because the transversely polarized $\rho$-meson is a massive $P$-wave state, its state can most generally be decomposed as the linear  combination of momentum basis states\footnote{This decomposition characterizes the $\rho$ production plane relative to the direction of $\v P$; the further decomposition we will discuss, of $\v P$ relative to the beam (quantization) axis, represents the decay plane of the $\rho$-meson.}
\begin{equation}
\label{eq:rho-Fock-decomposition}
    |\rho_{\v b}^{12}\rangle = \cos(\theta_{Pz}) \cos(\phi_{Pb}) |P_\parallel\rangle + \sin(\theta_{Pz})\cos(\phi_{Pb})
    |P_1^T\rangle + \cos(\theta_{Pz})\sin(\phi_{Pb})|P_2^T \rangle \,,
\end{equation}
where $|P_\parallel \rangle$ denotes the basis vector in the direction along $\v P$ and $|P_1^T\rangle$ and $|P_2^T\rangle$, the basis vectors in the directions orthogonal to it. Here $\theta_{Pz}$ denotes the polar angle of $\v P$ with respect to the $Z$ representing the beam direction, and $\phi_{Pb}$ is the azimuthal angle relative to the impact parameter in the YX-plane. These angles are illustrated in Fig.~\ref{fig:angle}. Note that in the STAR and ALICE experiments, the azimuthal angle has a $2\pi$-coverage around the beam axis.

It is convenient (and conventional) to choose the quantization axis of the $J=1$ P-wave state to lie along the beam direction, where as noted, the angle $\theta_{Pz}$ is the polar angle between the $Z$-axis and $\v P$. In the rest frame of the $\rho$-meson, where $\v p_1 = -\v p_2$, it is just  the polar angle of $\v p_1$ with respect to the $Z$-axis. This projection of the $\rho$-meson state can therefore be further expressed as the $M=0$ entangled state 
  \begin{equation}
   \label{eq:rho-longitudinal}
    \begin{split}
        |P_{\parallel} \rangle&= \cos(\theta_{Pz})\Big(|\pi^+(p_1)\pi^-(p_2)\rangle + |\pi^+(p_2)\pi^-(p_1)\rangle\Big) \,.
    \end{split}
\end{equation}
The rapidity of $\v P$ with respect to the beam is defined as $\eta= -\frac{1}{2}\ln(\tan(\theta_{Pz}))$. 

Note further that we have absorbed the $1/\sqrt{2}$ normalization of the two $M=0$ states with the coefficient of the $Y_1^0$ spherical harmonic\footnote{The $J=1$ spherical harmonics are $Y_1^{\pm 1} = \mp \sqrt{\frac{3}{8\pi}}\,\sin(\theta_{Pz})\,e^{\pm i\phi_{Pz}}$ for transverse polarizations and $Y_1^0=\sqrt{\frac{3}{4\pi}}\,\cos(\theta_{Pz})$ for longitudinal polarization.}, so that one can equivalently write, with the same normalization, the decomposition of the basis states in terms of the  $M=\pm 1$ spin states of the $\rho$-meson as 
\begin{equation}
\label{eq:rho-transverse}
    \begin{split}
        |P_T^1\rangle=
        -\sin(\theta_{Pz})\,e^{i\phi_{Pz}}|\pi^+(p_1)\pi^-(p_2)\rangle \qquad;\qquad |P_T^2\rangle=\sin(\theta_{Pz})\,e^{-i\phi_{Pz}}|\pi^+(p_2)\pi^-(p_1)\rangle\,.
    \end{split}
\end{equation}
where $\phi_{Pz}$ the azimuthal angle of $\v P$ in the lab frame. Substituting these in the first expression, we get 
\begin{eqnarray}
  |\rho_{\v b}^{12}\rangle &=&  \cos(\phi_{Pb}) \cos^2(\theta_{Pz})\Big(|\pi^+(p_1)\pi^-(p_2)\rangle + |\pi^+(p_2)\pi^-(p_1)\rangle\Big) \nonumber \\
  &-& \cos(\phi_{Pb})
  \sin^2(\theta_{Pz})\,e^{i\phi_{Pz}}|\pi^+(p_1)\pi^-(p_2)\rangle \nonumber\\ 
  &+& \frac{1}{2}\sin(\phi_{Pb})\sin(2\theta_{Pz})\,e^{-i\phi_{Pz}}|\pi^+(p_2)\pi^-(p_1)\rangle\,.
\end{eqnarray}

The amplitude then to produce $\pi^+$-meson with momentum $\v p_1$ or with momentum $\v p_2$ is obtained by projecting the Fock states $|\pi^+(p_1)\pi^-(p_2)\rangle$ and $|\pi^+(p_2)\pi^-(p_1)\rangle$ on the expression above, to give 
\begin{eqnarray}
\label{eq:M12}
    M_{12}^{\rho_{\v b}\rightarrow \pi^+\pi^-}+M_{12}^{\rho_{\v b}\rightarrow \pi^-\pi^+} &=& A(|\v P|,|\v q|) * \big(\left[ \cos(\phi_{Pb})\cos^2(\theta_{Pz}) -\cos(\phi_{Pb})
  \sin^2(\theta_{Pz})\,e^{i\phi_{Pz}}\right]  \nonumber\\
  &+& \left[\cos(\phi_{Pb})\cos^2(\theta_{Pz})+\frac{1}{2}\sin(\phi_{Pb})\sin(2\theta_{Pz})\,e^{-i\phi_{Pz}}\right]
  \big)\,.
\end{eqnarray}
Here $A$ is a function that characterizes the Breit-Wigner shape of the $\rho$-decay width and only depends on the modulus of $\v P$ and $\v q$. 

One can write an equivalent expression for $M_{21}$. For this, since one flips $\v b\rightarrow -\v b$ along the $Y$-axis, this is equivalent to replacing $\phi_{Pb} \rightarrow \pi - \phi_{Pb}$ everywhere in the above expression. This gives
\begin{eqnarray}
\label{eq:M21}
    M_{21}^{\rho_{\v b}\rightarrow \pi^+\pi^-}+M_{21}^{\rho_{\v b}\rightarrow \pi^-\pi^+}&=& A(|\v P|,|\v q|) * \big(\left[-\cos(\phi_{Pb})\cos^2(\theta_{Pz})+\cos(\phi_{Pb})
  \sin^2(\theta_{Pz})\,e^{i\phi_{Pz}}\right]  \nonumber\\
  &+& \left[-\cos(\phi_{Pb}) \cos^2(\theta_{Pz})- \frac{1}{2}\sin(\phi_{Pb})\sin(2\theta_{Pz})\,e^{-i\phi_{Pz}}\right]
  \big)\,.
\end{eqnarray}
Multiplying $M_{12}$ and $M_{21}$ respectively by the $ e^{\pm iq_\perp b_\perp \cos(\phi_{qb})}$ phase and adding the result (the relative $-$ sign is taken care of automatically due to the flip in the relative azimuthal angle), we obtain 
\begin{eqnarray}
\label{eq:M12+M21_1}
    M_{12}+M_{21} &=& A(|\v P|,|\v q|) \,2\,i\sin(q_\perp b_\perp\cos(\phi_{qb}))\nonumber \\
    &\times& \left[2\cos(\phi_{Pb})\cos^2(\theta_{Pz})-\cos(\phi_{Pb})
  \sin^2(\theta_{Pz})\,e^{i\phi_{Pz}}+\frac{1}{2}\sin(\phi_{Pb})\sin(2\theta_{Pz})\,e^{-i\phi_{Pz}}\right]\,.
\end{eqnarray}
This expression vanishes for $q_\perp\rightarrow 0$, as anticipated. Eq.~\eqref{eq:M12+M21_1} is our general result for the decay amplitude for the production of a $\pi^+\pi^-$ pair from the exclusive production of $\rho$-mesons in ultraperipheral collisions of identical ultrarelativistic nuclei. 

In the STAR experiment, these measurements are performed in a window of rapidity $\Delta \eta=\pm 1$ centered around $\eta=0$, corresponding to 
$\theta_{Pz}=\pi/2$. Our expression simplifies greatly for $\eta=0$, giving
\begin{eqnarray}
\label{eq:M12+M21_2}
    M_{12}+M_{21} = A(|\v P|,|\v q|) \,2\,(-i)\sin(q_\perp b_\perp\cos(\phi_{qb})) \cos(\phi_{Pq}+\phi_{qB})\,e^{i\phi_{Pz}}\,.
\end{eqnarray}
The $\cos(\phi_{Pb})$ dependence is what one expects from s-channel helicity conservation at the level of the amplitude~\cite{Bauer:1977iq};  we further replaced it here with $\phi_{Pb}=\phi_{Pq}+\phi_{qb}$. 
Taking the squared modulus of this expression, we first integrate over 
the azimuthal angle $\Phi_{Pz}$, which just gives a factor $2\pi$. We then integrate over $\phi_{qb}$ to obtain the azimuthal angle dependence on $\Phi_{Pq}$:
\begin{eqnarray}
\label{eq:squared-modulus}
   \int_0^{2\pi} d\phi_{qb} |M_{12}+M_{21}|^2 &\propto& \int_0^{2\pi} d\phi_{qb} \sin^2(q_\perp b_\perp\cos(\phi_{qb})) \cos^2(\phi_{Pq}+\phi_{qb})\nonumber \\
   &\propto& \int_0^{2\pi} d\phi_{qb} \left[1-\cos(2q_\perp b_\perp\cos(\phi_{qb}))\right]\cos^2(\phi_{Pq}+\phi_{qb})\,.
\end{eqnarray}
Note that much of the discussion in the literature has been in terms of spin-density matrices~\cite{Schilling:1969um,Bauer:1977iq,Ivanov:2004ax} which captures the relevant information at the level of the cross-sections. Expressing this information instead in terms of angular momentum states\footnote{In this context, see also a recent discussion in the context of DIS~\cite{Bhattacharya:2024sno}.} at the level of the amplitude helps clarify the entanglement inherent in the production of these exclusive states shedding further light into the underlying quantum dynamics. 

\section{$E^2 I^2$ from two particle resonant decays}
\label{section:3}
\subsection{Final analytical result and interpretation}
\label{section:3A}
We are now in a position to compute the cross-section for the two-particle distribution of $\pi^\pm$ pairs resulting from the resonant decay of $\rho$-mesons in UPCs by combining the results of the previous two subsections. We have two key interference effects in the cross-section. One is the interference due to the phase factor related to the switch $A_1 \leftrightarrow A_2$. This effect is responsible for the destructive interference of the amplitudes as seen for $q_\perp\rightarrow 0$ in Eq.~(\ref{eq:M12+M21_1}). The other effect is the phase information in the $\rho\rightarrow \pi^+\pi^-$ decay which accounts for the spin-$1$ of the $\rho$ being transmitted to the angular momentum of the entangled spin-$0$ pions. The latter, as noted, is what is responsible for $E^2 I^2$ and leads to the azimuthal cosine angle orientations seen in Eq.~(\ref{eq:M12+M21_1}) between the decay pions and the impact parameter plane.

The specific interferometric measurement we will consider is the average two particle multiplicity corresponding to the inclusive probability~\cite{Itzykson:1980rh,Gelis:2001dh} for exclusive production of the $\pi^+\pi^-$-pair from $\rho$-decays. For the coherent case, where the nuclei remain intact in the scattering, this is given by
\begin{equation}
\label{eq:interference1}
    \frac{d^2 N^{\rm coh.}}{dP_\perp^2 dq_\perp^2 d\phi_{\v P_\perp \v q_\perp}}=\int d\v b_\perp^2 d\phi_{qb} \;|M_{12\rightarrow \pi^+\pi^-} + M_{21\rightarrow \pi^+\pi^-}|^2\,.
\end{equation}
Substituting Eqs.\eqref{eq:final-12-rho-amplitude},  and \eqref{eq:M12+M21_2} in the above expression, and employing Eq.~\eqref{eq:squared-modulus}, we obtain 
\begin{eqnarray}
\label{eq:interference2}
    \frac{d^2 N^{\rm coh.}}{dP_\perp^2 dq_\perp^2 d\theta_{\v P_\perp \v q_\perp}d^2{\v \Delta K_\perp}}&=&C\,|A(|\v P|,|\v q|)|^2\,\int d\v b_\perp^2 F_{\rm QED}^2\Big(\v q_\perp - \frac{1}{|\v b-\v R|}\Big)\nonumber \\
    &\times&
      \int \frac{d^2\v K_\perp}{(2\pi)^2} \, \Bigg |M_{\gamma {\mathbb P}\rightarrow\rho}^{T\lambda_{\mathbb P}\rightarrow T}(\v q_\perp-\v \Delta K_\perp, \v K_\perp,\v q_\perp)\Bigg |^2\, \Bigg |\int_{|\v X|<R}\,d^2 \v X\,e^{i\v \Delta K_\perp\cdot \v X}\,P(\v X, \v K_\perp)\Bigg |^2\nonumber \\
&\times& \int_0^{2\pi} d\phi_{qb} \Big[1-\cos(2q_\perp b_\perp\cos(\phi_{qb}))\Big]\cos^2(\phi_{Pq}+\phi_{qb})\,.
\end{eqnarray}
Here $C$ is an overall constant absorbing various other constants in the computation. Further, $|A(|\v P|,|\v q|)|^2$ is the invariant decay probability for 
$\rho\rightarrow \pi^+ \pi^-$. This is typically given by a Breit-Wigner distribution where the parameters are the $\rho$-mass, its decay width, and the $\rho\rightarrow \pi\pi$ decay constant. Performing the angular integration over $\phi_{qb}$ in Eq.~(\ref{eq:interference2}), we obtain 
\begin{equation}
    \begin{split}
         2\pi \Big[\Big(\frac{1}{2}-\frac{J_1(2q_\perp b_\perp)}{2q_\perp b_\perp}+\frac{J_2(2q_\perp b_\perp)}{2}  \Big)+ \frac{J_2(2q_\perp b_\perp)}{2}  \cos(2\phi_{Pq})\Big]\,.
    \end{split}
\end{equation}
Our final result for the coherent exclusive cross-section  $A_1\,A_2\rightarrow \pi^+\pi^-$ can be expressed as 
\begin{eqnarray}
\label{eq:interference-final}
 \frac{d^2 N^{\rm coh.}}{dP_\perp^2 dq_\perp^2 d\theta_{\v P_\perp \v q_\perp}d^2{\v \Delta K_\perp}}  &=&{\bar C}\,|A(|\v P|,|\v q|)|^2\,\int d\v b_\perp^2 F_{\rm QED}^2\Big(\v q_\perp - \frac{1}{|\v b-\v R|}\Big)
       \nonumber \\
    &\times&
\int \frac{d^2\v K_\perp}{(2\pi)^2} \, \Bigg |M_{\gamma {\mathbb P}\rightarrow\rho}^{T\lambda_{\mathbb P}\rightarrow T}(\v q_\perp-\v \Delta K_\perp, \v K_\perp,\v q_\perp)\Bigg |^2\, \Bigg |\int_{|\v X|<R}\,d^2 \v X\,e^{i\v \Delta K_\perp\cdot \v X}\,P(\v X, \v K_\perp)\Bigg |^2\nonumber \\
&\times&  \Big[\Big(\frac{1}{2}-\frac{J_1(2q_\perp b_\perp)}{2q_\perp b_\perp}+\frac{J_2(2q_\perp b_\perp)}{2} \Big)+ \frac{J_2(2q_\perp b_\perp)}{2}  \cos(2\phi_{Pq})\Big]\,,
\end{eqnarray}
where we have now absorbed a factor $2\pi$ in the constant 
${\bar C}$. For fixed $q_\perp$, this result clearly has a $\cos(2\phi_{Pq})$ modulation. As noted earlier, this signal is the combined result of the interference from the interchange of nucleus $1\leftrightarrow 2$, and from $E^2 I^2$ in the amplitude for $\rho\rightarrow \pi^+\pi^-$. The modulation is also seen in the STAR and ALICE UPC data\footnote{The STAR data is consistent with higher harmonic contributions as well, though they are small.} though, as we shall soon discuss at length, the incoherent cross-section needs to be computed carefully before one concludes that the modulation seen in the data is due to the model-independent $E^2 I^2$ effect rather than a dynamical effect in a theory framework or a combination of the two.

\begin{figure}[H]
        \centering
        \includegraphics[width=0.5\linewidth]{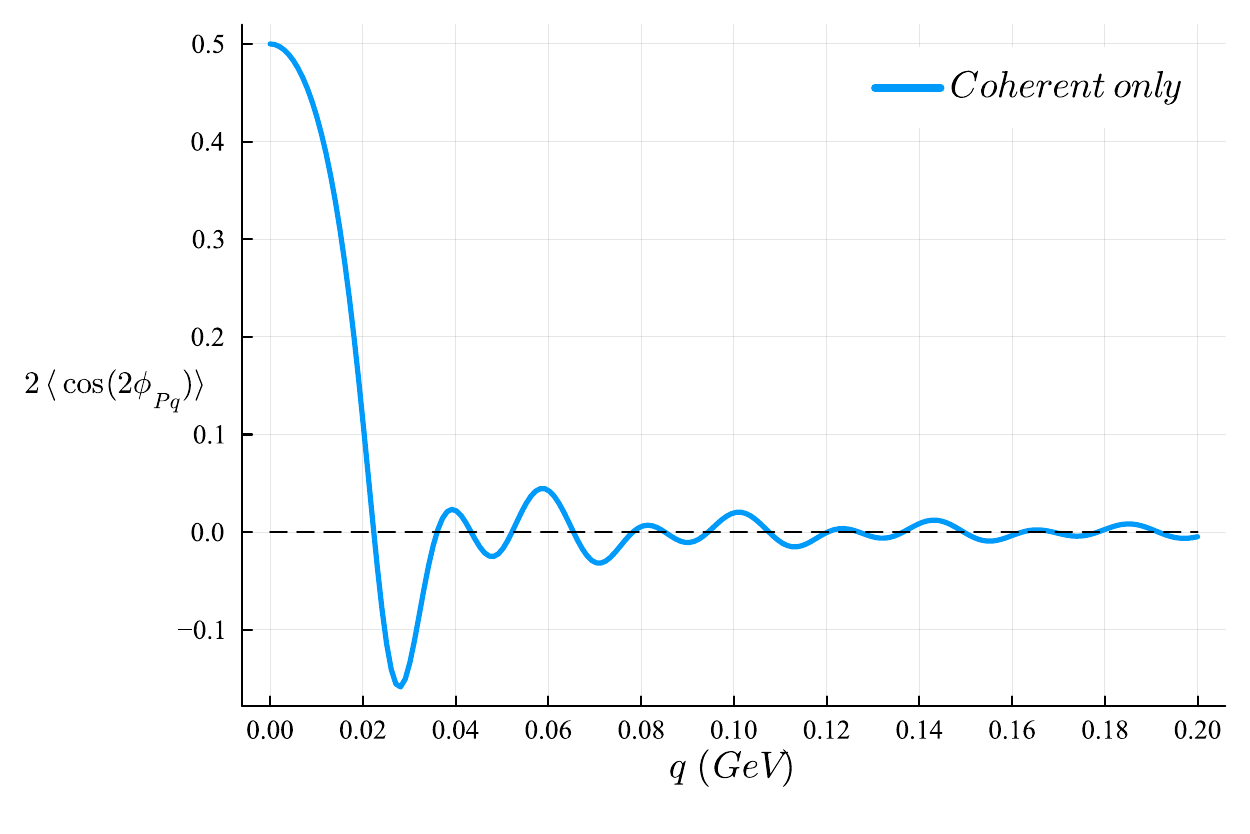}
    \caption{Illustration of  result in Eq.~\eqref{eq:v2} as a function of $q_\perp$, with 
    input from the coherent cross-section in Eq.~\eqref{eq:interference-final}. }
    \label{fig:coherent}
\end{figure}

The normalized anisotropy coefficient of the $\cos(2\phi_{Pq})$ modulation of this result has the advantage that it is insensitive to the overall normalization factors. To extract it, we define 
\begin{equation}
    V_n(q_\perp,b_{\rm min})=\int P_\perp d  P_\perp \int_0^{2\pi} d \phi_{Pq} \;   \frac{d^2 N^{\rm coh.}}{dP_\perp^2 dq_\perp^2 d\phi_{\v P_\perp \v q_\perp}} \cos(n \phi_{Pq})\,,
\end{equation}
with the $\cos(n\phi_{Pq})$ anisotropy expressed as
\begin{equation}
\label{eq:v2}
    \begin{split}
        \langle \cos n \phi_{Pq}) \rangle(q_\perp,b_{\rm min}) =\frac{1}{2} \frac{V_n(|\v q|)}{V_0(|\v q|)}\,.
    \end{split}
\end{equation}
An illustrative result for $\langle \cos (2 \phi_{Pq}) \rangle$ is shown in Fig.~\ref{fig:coherent}.  In obtaining this result, we integrated over the impact parameter in Eq.~\eqref{eq:interference-final}, employing a standard expression for the photon form factor~\cite{Vidovic:1992ik}.
The result is sensitive to $b_{\rm min}$ and less so to $b_{\rm max}$; for this case, we took $b_{\rm min}=15 \;\rm fm$, and $b_{\rm max}=30 \;\rm fm$. At large $q_\perp$, the behavior of the ratio in Fig.~\ref{fig:coherent} goes to zero as can be anticipated from the asymptotics of the Bessel functions in Eq.~\ref{eq:interference-final}.
We observe further that the anisotropy goes to a constant value as $q_\perp\rightarrow 0$, despite the  overall cross-section going to zero as $q_\perp\rightarrow 0$, as mentioned previously. This is because the two terms in the ratio $\frac{1}{2} \frac{V_2(|\v q|)}{V_0(|\v q|)}$ have the same functional form in this limit. Thus if there is an arbitrarily small but additional contribution that is finite when $q_\perp\rightarrow 0$, then $\langle \cos (n \phi_{Pq}) \rangle$ will go to zero as $q_\perp\rightarrow 0$.

This is indeed likely the case in the UPC experiments since what is measured is not the coherent exclusive cross-section alone but the sum over the coherent+incoherent contributions. 
While this complicates the interpretation of the results, the incoherent contribution is interesting in its own right as we will now discuss.

\subsection{Total inclusive distribution for exclusive $A_1\,A_2\rightarrow \rho\rightarrow \pi^+\pi^-$ }
\label{subsection:3B}
In the discussion thus far, we assumed that the exclusive production of $\rho$-mesons is in {\it coherent} UPC collisions, where the nuclei remain intact. The total cross-section for such exclusive processes however also includes an {\it incoherent} component corresponding to the break up of the nuclei into nucleons~\cite{Frankfurt:2022jns}. In these  events, the rapidity gap from the nucleus to the vector meson is still preserved, as indicated by the absence of the emission of additional pions not coming from the resonant decay. 

\begin{figure}[H]
\centering
\includegraphics[width=0.5\linewidth]{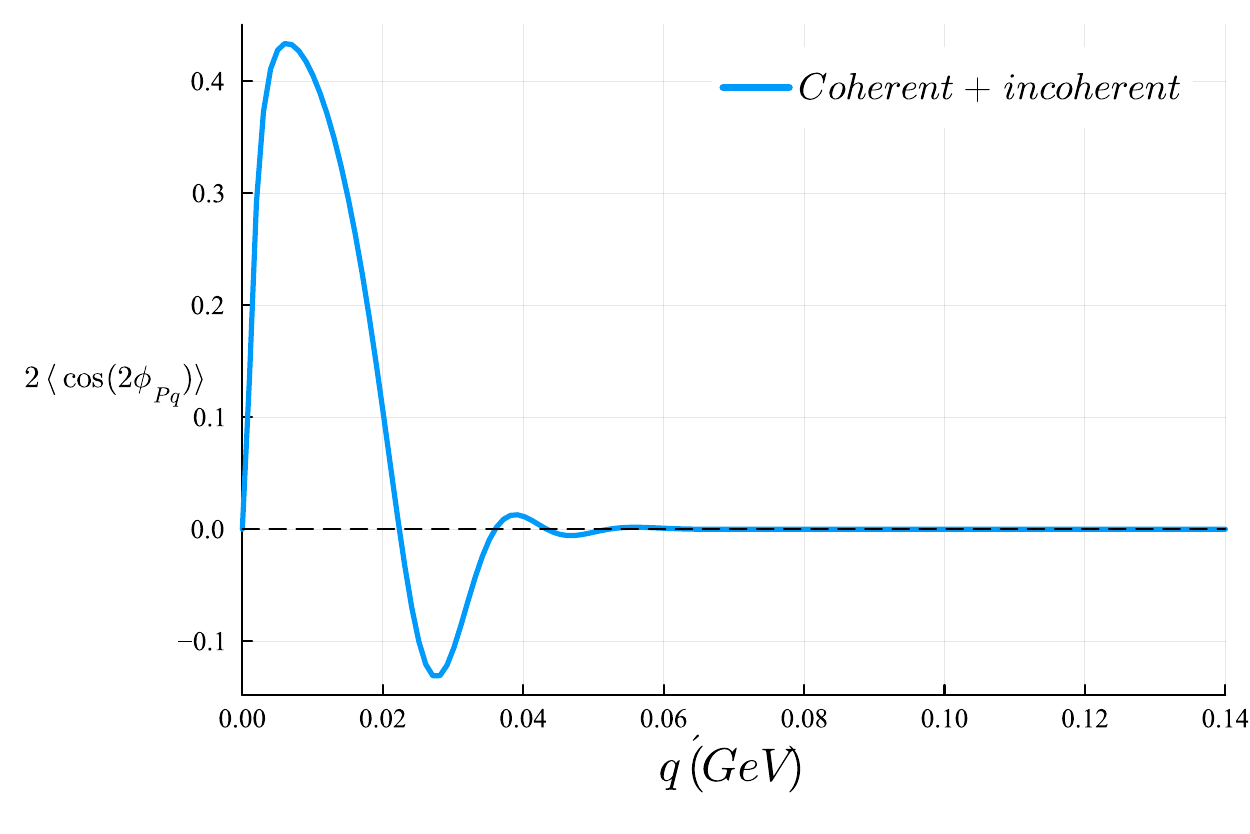}
    \caption{Illustration of the result in Eq.~\eqref{eq:v2} as a function of $q_\perp$ after adding the simple model of the incoherent contribution in Eq.~\eqref{eq:incoherent-toy} to the expression in Eq.~\eqref{eq:coherent-toy}. The incoherent piece in this toy model only contributes to $V_0$.}
    \label{fig:coherent+incoherent}
\end{figure}

In this subsection, we will first use a simple toy model of the incoherent exclusive cross-section to examine how this changes the coherent result shown in Fig.~\ref{fig:coherent}. 
Adding the incoherent contribution, the anisotropy coefficients in our framework can be simply expressed as 
\begin{equation}
\label{eq:Vn}
    V_n(q_\perp,b_{\rm min})=\int P_\perp d  P_\perp \int_0^{2\pi} d \phi_{Pq} \;   \Bigg(\frac{d^2 N_{\rm toy}^{\rm coh.}}{dP_\perp^2 dq_\perp^2 d\phi_{\v P_\perp \v q_\perp}} + \frac{d^2 N_{\rm toy}^{\rm incoh.}}{dP_\perp^2 dq_\perp^2 d\phi_{\v P_\perp \v q_\perp}}\Bigg) \cos(n \phi_{Pq})\,.
\end{equation}
For the simplified model, we take 
\begin{eqnarray}
\label{eq:coherent-toy}
    \frac{d^2 N_{\rm toy}^{\rm coh.}}{dP_\perp^2 dq_\perp^2 d\phi_{\v P_\perp \v q_\perp}} &=& A_c\,\exp(- R_0^2\, q_\perp^2)\,
    \int d\v b_\perp^2 F_{\rm QED}^2\Big(\v q_\perp - \frac{1}{|\v b-\v R|}\Big)\,\,\nonumber \\
   &\times&  \Big[\Big(\frac{1}{2}-\frac{J_1(2q_\perp b_\perp)}{2q_\perp b_\perp}+\frac{J_2(2q_\perp b_\perp)}{2} \Big)+ \frac{J_2(2q_\perp b_\perp)}{2}  \cos(2\phi_{Pq})\Big]\,,
\end{eqnarray}
and 
\begin{equation}
\label{eq:incoherent-toy}
    \frac{d^2 N_{\rm toy}^{\rm incoh.}}{dP_\perp^2 dq_\perp^2 d\phi_{\v P_\perp \v q_\perp}}= \frac{A_i/Q_0^2}{(1+q_\perp^2/Q_0^2)^2}\,.
\end{equation}
Eq.~\eqref{eq:coherent-toy} is a much simplified version of the general expression in Eq.~\eqref{eq:interference-final}, where we have replaced all the complexity of the coherent $\gamma{\mathbb P}\rightarrow \rho$ cross-section with $A_c\,\exp(-R_0^2\, q_\perp^2)$. Here $R_0=6.72\pm 0.02$ fm is the radius of a gold nucleus. 
The incoherent contribution in Eq.~(\ref{eq:incoherent-toy}) is parametrized by a dipole form factor with $Q_0^2=0.099$ GeV$^{2}$ and $A_i$ is a dimensionful constant. Clearly, in this parametrization, there is a finite incoherent component even as $q_\perp\rightarrow 0$. As we noted in the previous subsection, the presence of such a component is responsible for the dip in $2\langle\cos(2\phi_{Pq})\rangle$, as $q_\perp\rightarrow 0$, shown in Fig.~\ref{fig:coherent+incoherent}.
The results for $ V_2(q_\perp,b_{\rm min})$ are shown in Fig.~\ref{fig:coherent+incoherent} for different fractions of $A_c$ and $A_i$. 

\begin{figure}[H]
        \centering
\includegraphics[scale=0.5]{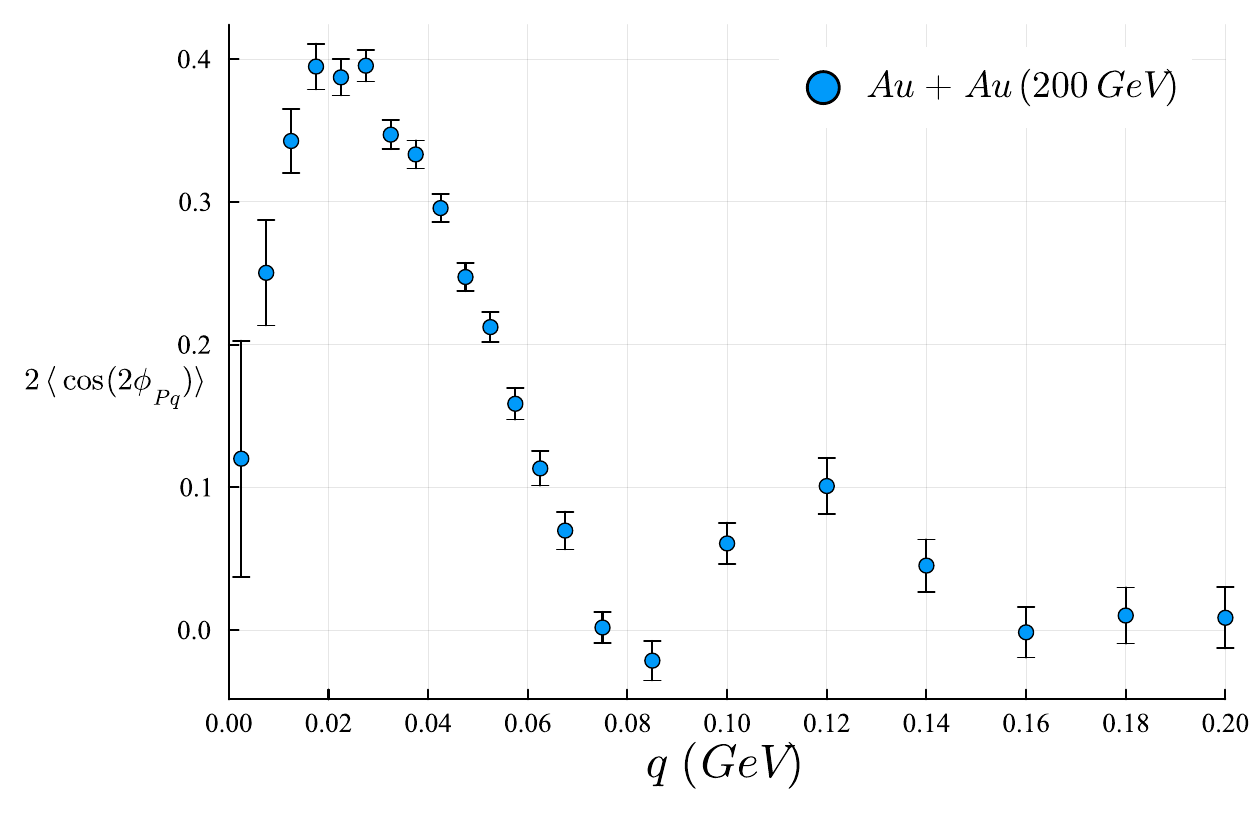}
    \caption{
   Data from the STAR collaboration \cite{STAR:2022wfe} for ``Au+Au'' UPCs at center of mass energy $\sqrt s=200\, \rm GeV$.}
    \label{fig:data-coherent+incoherent}
\end{figure}

Though the result shown in Fig.~\ref{fig:coherent+incoherent} is oversimplified, it is useful to qualitatively 
compare its features to the UPC data to understand better what features of the data are not captured by the  expressions in Eqs.~(\ref{eq:coherent-toy}) and (\ref{eq:incoherent-toy}). The data from the STAR experiment are shown in Fig.~\ref{fig:data-coherent+incoherent}. Firstly, we observe that the rich systematics seen in the 
data is occuring for $q_\perp < 200$ MeV. At these values of $q_\perp$, we do not expect perturbative, or more generally, weak coupling methods to be useful. Comparing Fig.~\ref{fig:coherent+incoherent} to the data more closely, we see that both data and model show a rapid decrease towards $q_\perp\rightarrow 0$. The first peak in the model is $q_\perp\sim 10$ MeV but then rapidly decreases with increasing $q_\perp$, to have a minimum at $q_\perp\sim 20$ MeV. In contrast, the data is much broader, with the minimum at $\sim 80$ MeV instead. The behavior at larger $q_\perp$ disagrees completely with the toy model. 

The STAR collaboration has performed fits to their data with different functional forms~\cite{STAR:2022wfe}. 
According to the STAR fits, the coherent exclusive cross-section dominates at very low $q_\perp\sim 10-20$ MeV, with the incoherent cross-section increasing, providing a comparable contribution by $q_\perp\sim 100$ MeV, and   dominating the coherent cross-section fully by $q_\perp\sim 200$ MeV. 

The qualitative comparison above indicates that a) a more detailed computation of Eq.~(\ref{eq:interference-final}) relative to the simplified model in Eq.~(\ref{eq:coherent-toy}) may be necessary, but more importantly, 
b) the incoherent cross-section can be computed in this framework also applying the $E^2 I^2$ framework. We will outline the key elements of the computation below but leave the more detailed study and the possible use of this study for model-independent extraction of pomeron properties to future work.

A systematic computation of the total exclusive cross-section can be performed in the framework developed here to reveal the rich interplay between the coherent and incoherent contributions. We will first outline the general structure of this computation and then discuss the implications for the azimuthal anisotropy. On the surface, both coherent and incoherent contributions appear to have the same angular dependence. A closer examination of our result will reveal this not to be the case.  The path to identifying the incoherent component is to first compute the total exclusive vector meson cross section, and to subtract from it the coherent component. This is determined by first squaring the color averaged amplitude in a nucleus and then averaging over nucleon positions/color singlet patches\footnote{The formula should be interpreted as covering the four possibilities: i) the sum $\langle M_{12}\rangle_{\rho,N}+\langle M_{21}\rangle_{\rho,N}$ of the averaged amplitudes in color singlet patches of both nuclei, each then averaged over nucleon position in the respective nuclei, with the sum then squared. This corresponds to the no-break up configuration. ii) The sum $\langle M_{12}\rangle_{\rho,N} + \langle M_{21}\rangle_{\rho}$ with both color singlet and nucleon averages performed at the level of the amplitude in nucleus 2, but the average over nucleon positions in nucleus 1 performed only at the level of the squared amplitude. This corresponds to the case where nucleus 1 breaks up but not nucleus 2. iii) The reverse configuration for the two nuclei. iv) The ``double break-up" case, where the average over nucleon positions performed for both $M_{12}$ and $M_{21}$ at the level of the cross-section. 
}:
\begin{equation}
\label{eq:incoherent-multiplicity}
    \frac{d^2 N^{\rm coh.+incoh.}}{dP_\perp^2 dq_\perp^2 d\theta_{\v P_\perp \v q_\perp}}\propto  |\langle M_{12\rightarrow \rho}^{T\lambda_{\rho_{\v b}}}(\v q,\v b)+ M_{21\rightarrow \rho}^{T \lambda_{\rho_{\v b}}}(\v q,\v b)\rangle_{\rho_{\rm QCD}}|_{\rm nucleon}^2\times
    |M_{\rho_{\v b}\rightarrow \pi^+\pi^-}|^2\,.
\end{equation}
The total exclusive $\rho^0$-meson distribution  can be straightforwardly deduced from Eq.~\eqref{eq:finalB-12-rho-amplitude}:
\begin{eqnarray}
\label{eq:total-cross-section}
 \frac{d^2 N^{\rm coh.+incoh.}}{dP_\perp^2 dq_\perp^2 d\theta_{\v P_\perp \v q_\perp}d^2{\v \Delta K_\perp}}    &=& {\bar C_{\rm tot.}}\,|A(|\v P|,|\v q|)|^2\,\int d\v b_\perp^2 F_{\rm QED}^2\Big(\v q_\perp - \frac{1}{|\v b-\v R|}\Big)\nonumber\\
       &\times& \int \frac{d^2\v K_\perp}{(2\pi)^2} \, \int_{|\v X|<R}\,d^2 \v X\,e^{i\v \Delta K_\perp\cdot \v X}\,P^2(\v X, \v K_\perp)\,\Bigg |M_{\gamma {\mathbb P}\rightarrow\rho}^{T\lambda_{\mathbb P}\rightarrow T}(\v q_\perp-\v \Delta K_\perp, \v K_\perp,\v q_\perp )\Bigg |^2\nonumber \\
&\times&  \Big[\Big(\frac{1}{2}-\frac{J_1(2q_\perp b_\perp)}{2q_\perp b_\perp}+\frac{J_2(2q_\perp b_\perp)}{2} \Big)+ \frac{J_2(2q_\perp b_\perp)}{2}  \cos(2\phi_{Pq})\Big]\,.
\end{eqnarray}
Here ${\bar C_{\rm tot.}}$ is an overall constant factor. 

Comments regarding this result are in order. Comparing this result to Eq.~\eqref{eq:interference-final}, we observe that the incoherent cross-section is proportional to 
\begin{equation}
    \langle P^2(\v X, \v K_\perp)\rangle_{\v X} - \langle P(\v X, \v K_\perp)\rangle_{\v X}^2 \,.
\end{equation}
The incoherent cross-section is therefore sensitive to fluctuations in the pomeron distribution in a nucleus, a result well-known in the literature~\cite{Miettinen:1978jb,Kovchegov:1999kx,Kowalski:2008sa}. For a discussion of subtleties in the interpretation of the result, in particular, the somewhat ambiguous separation of coherent and incoherent cross-section when additional photons are exchanged/emitted, see \cite{Klein:2023zlf}. Such contributions may be important for a quantitative analysis of the UPC data. A key point from our more limited perspective is that the average over the color singlet patches of the pomeron distribution or its square does not necessarily correlate directly with the nuclear break-up which can occur due to electromagnetically induced processes. 

The other noteworthy point relates to our earlier discussion on the angular dependence. Eq.~\eqref{eq:total-cross-section} has an identical sensitivity (on the surface) to the angular $\pi^+\pi^-$ distribution, albeit weighted differently. This is qualitatively different from Eq.~\eqref{eq:incoherent-toy}. If true, it poses a problem with respect to the UPC data. The coherent contribution to the azimuthal anisotropy $V_2$, as noted earlier, clearly misses qualitative features of the data shown in \cite{STAR:2022wfe}. Adding an incoherent contribution with the qualitatively similar behavior seen in Eq.~\eqref{eq:total-cross-section} will not alter this result.

The way forward in resolving this issue follows from a closer look at Eq.~\eqref{eq:total-cross-section}. Both the l.h.s and r.h.s depends on $\v \Delta K_\perp$, which is the momentum recoil suffered by the nucleon/patch $\v X$ inside the nucleus. From Eq.~\eqref{eq:finalA-12-rho-amplitude}, we see that it satisfies the $\delta$-function constraint $\v \Delta K_\perp = \v q_\perp- \v k_\perp$. For low values of $\v q_\perp$ around the first peak, its momentum is comparable to the photon transverse momentum $k_\perp\sim 1/b_{\rm min.}\sim 15$ MeV. However when $\v q_\perp > k_\perp$, the delta-function constraint gives 
$\v q_\perp \approx \v \Delta K_\perp$. Thus the $\rho$-meson momentum $\v q_\perp$ becomes lockstep with the recoil momentum $-\Delta K_\perp$ of the nucleon/color singlet patch $\v X$ in the nucleus. In other words, the correlation between the recoil vector and the $\rho$-meson momentum also correlates the production and decay planes in the resonant exclusive cross-section. 

Since $\v \Delta K_\perp$ is not measured in the UPC experiments, the results average over the angular integral in $\v \Delta K_\perp$ when integrating over $\v \Delta K_\perp$ in Eq.~\eqref{eq:total-cross-section}. One is thus effectively smearing over the angular correlation between $\v P_\perp$ and $\v q_\perp$ gradually as $\v q_\perp > k_\perp$. While this is an important systematic effect that must be taken into account in an {\it ab initio} computation, it requires a more detailed computation of the $\gamma \mathbb P\rightarrow \rho$ matrix element. Work in this direction will be reported separately. 

\subsection{Other exclusive final states}
\label{subsection:3C}

The application of the $E^2 I^2$ analysis to extract fundamental nonperturbative information from UPC data requires a significant extension of our study to compute the incoherent exclusive cross-section. 
With this important necessary analysis,  the general $E^2 I^2$ framework for the exclusive resonant production of $\pi^+\pi^-$ pairs in UPCs can be applied to other similar final states such as 
$\phi\rightarrow K^+ K^-$ and $J/\psi\rightarrow e^+ e^-$ to see if a global analysis can help extract model-independent nonperturbative information.
With regard to heavy Onium states, the intermediate ``CGC motivated" steps in the computation in section~\ref{section:2B} are relevant. Specifically, the $
\gamma \mathbb P\rightarrow Q\bar Q$ short distance amplitude can be computed in the CGC EFT and the long distance matrix elements $Q\bar Q\rightarrow J/\psi, \Upsilon,\cdots$ determined using NRQCD~\cite{Kang:2013hta}. The polarization of $J/\psi$ produced in {\it inclusive} proton-proton and proton-nucleus collisions has been studied in this framework by reconstructing the angular distribution of lepton pairs, albeit employing the spin density matrix formalism~\cite{Ma:2018qvc,Stebel:2021bbn}; applications to $J/\psi$ polarization in semi-inclusive DIS have been considered previously in \cite{DAlesio:2021yws}. Because these are computations for inclusive final states, the process is not an illustration of $E^2 I^2$; the extension of this framework to the exclusive case in UPC's is straightforward. A striking result by the ALICE collaboration is that SCHC is also obeyed for $J/\psi$ polarization in UPC collisions~\cite{ALICE:2023svbb}; while there is considerable discussion in the literature, a complete theoretical understanding remains elusive. 

As we noted earlier, a similar analysis can be performed for $C=1$ vector meson exclusive decays, allowing access to the nuclear odderon distribution. This follows from the observation in \cite{Dumitru:2018vpr} that the structure of the correlator in Eq.~\eqref{eq:color-scattering-amplitude} can in general be expressed as 
\begin{equation}
    P(\v X,\v K) + i\, O(\v X, \v  K)\,.
\end{equation}
The pomeron piece contributes to $C=-1$ exclusive vector meson decays while the second term, the odderon operator, contributes to $C=+1$ final states. The rest of the computation for the latter case goes through identically as discussed here. A key channel to search for is the production of $\chi_c$ mesons in UPC's. This channel has been considered previously in the context of odderon discovery for the EIC~\cite{Benic:2024pqe}, in particular with reference to isolating this channel from Primakoff process arising from photon exchange. An $E^2 I^2$ analysis of $\chi_c$ decays can help distinguish between the two mechanisms. 

A further interesting possibility in exclusive few-particle decays of vector mesons in UPC's is the process $J/\psi\rightarrow \Lambda\bar{\Lambda}$. The $J/\psi$ has $J^P=1^-$, while the $\Lambda\bar{\Lambda}$ hyperon-pair is expected to be in a spin triplet state arising from the two spin $1/2$ particles. The interesting twist here is the weak decay of the $\Lambda$ hyperons provides clean information on the entangled state. 
Indeed, this process has long been seen as an attractive high energy process~\cite{Tornqvist:1980af,Hao:2009kj,Chen:2013epa,Qian:2020ini,Gong:2021bcp,Barata:2023jgd,Wu:2024mtj,Fabbrichesi:2024rec} for tests of the CHSH inequality~\cite{Clauser:1969ny}. 

\subsection{$E^2 I^2$ in context of recent literature on exclusive $\rho$ decays in UPC's}
\label{subsection:3D}

The study of photoproduction of vector mesons has a long history. A discussion of the spin density matrix formalism in this context can be found in \cite{Schilling:1969um,Schilling:1973ag}. Early work in theory and experiment, including a discussion of SCHC and other mechanisms, is summarized in \cite{Bauer:1977iq}. A comprehensive discussion of exclusive vector meson production in DIS, in particular of HERA data, is given in \cite{Ivanov:2004ax}. Exclusive vector meson production, within the broader context of diffractive final states, has been reviewed recently in \cite{Frankfurt:2022jns}. 

With regard to vector meson production in UPCs, there is already a significant literature. The pioneering work in \cite{Klein:1999gv,Klein:2002gc} brought to the fore the role of entanglement in exclusive decays in UPCs. The introduction of the polarization and its correlation with the impact parameter predicted the azimuthal angular distribution as a function of $|t|$ and resolved the puzzle of the $|t|$ distribution observed in experiments~\cite{Zha:2020cst,Xing:2020hwh,STAR:2022wfe}. Subsequent theory work can be broadly classified into vector meson dominance initiated Glauber/Gribov-type scattering models~\cite{Guzey:2022qvc,Zha:2020cst} and color dipole models~\cite{Bendova:2020hbb,Goncalves:2020vdp,Goncalves:2022ret,Xing:2020hwh,Hagiwara:2020juc,Mantysaari:2022sux,Mantysaari:2023prg}. Discussions in the framework of the latter set of models either implicitly or explicitly fall within the framework of the CGC EFT.  In particular, developing earlier work on exclusive photoproduction~\cite{Zha:2020cst,Hagiwara:2021xkf,Xing:2020hwh}, a careful and detailed comparison to the STAR data was performed in this framework in \cite{Xing:2020hwh,Mantysaari:2023prg}. 

Situating our work within this body of work will be helpful for further progress in the analysis of UPCs. Firstly, as noted earlier, while our computation was initially formulated in the CGC EFT framework, we were able to express the result for the $\rho$-production amplitude in a model-independent way, as shown for instance in 
Eq.~\eqref{eq:final-12-rho-amplitude}, as a convolution of the photon and pomeron flux factors, the matrix element for $\gamma{\mathbb P}\rightarrow q\bar q$, and that for $q\bar q\rightarrow \rho$. This can however be combined into an overall $\gamma \mathbb P\rightarrow \rho$ matrix element since it's not clear that the dipole model framework is valid at the non-perturbative scales relevant for $\rho$
-meson production in UPC's. Notably, the vector meson shadowing models~\cite{Guzey:2022qvc,Zha:2020cst} can also be accommodated within the framework. One can therefore focus on what data reveals about pomeron spin couplings, shadowing, and energy evolution. 

A further aspect of our work that has general validity and had not been noted previously is the observation that the correlation between the $\rho$-meson momentum and nucleon recoil with increasing transverse momentum induces a correlation between the production and decay planes, influencing the angular distribution of pair momenta. In CGC-based models, fits are performed to HERA exclusive vector meson data to extract the nucleon $t$ (or equivalently, the $-\v \Delta K_\perp$) dependence~\cite{Kowalski:2003hm,Rezaeian:2012ji,Schenke:2012wb,Mantysaari:2016ykx}. Its not clear however that this framework is applicable in region where $q_\perp\leq 0.2$ GeV. Once our computation of incoherent exclusive cross-sections is complete, a careful comparison of the model-independent features of our computation at finite $t$ to the high precision UPC data has the potential to reveal novel features of pomeron dynamics. 

The final distinguishing feature of our work is $E^2 I^2$ aspects of resonant few-particle decays. This is not emphasized in much of the recent literature, with most discussions traced back to the seminal papers on the spin density matrix formalism in the 1970's~\cite{Schilling:1969um,Schilling:1973ag,Bauer:1977iq}. The rest of the discussion is rooted in perturbative QCD computations of angular distributions; it is fair to ask whether these latter studies are relevant momentum scales less than $0.2$ GeV. Our study suggests that the $E^2 I^2$ formulation of the problem at amplitude level may allow greater insight into entangled states otherwise obscured by ``black box" applications of the spin density matrix formalism. In this case, general features of entangled states can allow us to uncover key phase information that may provide fresh insight into strong interaction dynamics. 

\section{Summary and outlook}
\label{section:4}
Hanbury-Brown--Twiss interferometry is now a widely used tool in diverse fields from astronomy, to heavy-ion physics, to quantum optics and cold atom physics. In its conventional usage, it relies on the coincidence measurements of indistinguishable quantum states. A novel idea by Cotler and Wilczek demonstrates that the HBT interferometric signal can be recovered for distinguishable quantum states by first entangling them via a unitary transformation and then employing a filter to project out the reduced density matrix of a subset of states. 

We introduced in this work a variant of this Entanglement Enabled Intensity Interferometry ($E^2 I^2$) idea  and discussed its application in the analysis of data in resonant exclusive decays of vector mesons. A particularly clean illustration is provided by  measurements in ultrarelativistic ultraperipheral nuclear collisions at RHIC and the LHC. We constructed a theoretical framework with minimal model assumptions that allowed us to compute $E^2 I^2$ correlations of $\pi^+\pi^-$ pairs in the resonant decay of $\rho$-mesons in UPCs. Our results for the coherent exclusive cross-section generalize straightforwardly to other exclusive vector meson decays. The interferometric correlations are sensitive to pomeron distributions and fluctuations in nuclei. They have the potential to provide pomeron spin and spin-flip couplings and provide novel input into searches for the predicted $C=-1$ color singlet odderon vacuum exchange partner of the $C=+1$ pomeron. 

An important step in fully assessing the utility of $E^2 I^2$ is to perform a quantitative comparison of our general framework with UPC data employing pomeron models with differing assumptions, and likewise, differing $\gamma \mathbb P\rightarrow \rho,\phi,J/\psi,\cdots$ non-perturbative matrix elements. As we discussed, a qualitative comparison with data shows significant disagreement in the regime where incoherent contributions are becoming important, and especially where they dominate. Work is in progress in 
computing the incoherent exclusive cross-section; within the $E^2 I^2$ framework, this analysis contains nontrivial features that have not been included previously. This work in progress will be reported separately. If this describes qualitative features of the data, enabling one to extract model-independent pomeron dynamics from a subsequent quantitative comparison, there are several directions to pursue. One is to reexamine what one can learn about the structure of different nuclei from exclusive vector meson final states. 
There have been several attempts to apply UPC exclusive vector meson studies on different nuclei to uncover features of their nuclear structure; for a general review, see \cite{Bally:2022vgo}. However as discussed, to reliably apply UPC studies towards this goal, one needs to first develop a better physically motivated understanding of the separation of coherent and incoherent contributions to exclusive vector meson production. 

Since the purpose of  $E^2 I^2$ approach is to identify entangled final states at the amplitude level and the information they provide on the strongly interacting dynamics of confining pomeron and saturated gluon configurations, the analysis techniques identified can be extended from UPC studies to deeply inelastic scattering. We anticipate $E^2 I^2$ should be especially valuable at the Electron-Ion Collider where one has more control over the initial scattering states relative to UPC's.

An intriguing possibility is to extend the $E^2 I^2$ ideas outlined here for resonant bound states to understand the spacetime structure and correlations in expanding clouds of mesoscopic ultracold 2-D Fermi gases released from an optical trap~\cite{Brandstetter:2023jsy}. 
Such ``table top" experiments are feasible and represent a promising direction for future research\footnote{We thank Maciej  Ga{\l}ka and Selim Jochim for useful discussions on this topic.}.  Not least, we note that interferometry and quantum computing have a rich ``entangled history"~\cite{Jin:2023gcv}. It is therefore natural to explore whether one can construct model Hamiltonians of the underlying hadronization process\footnote{An example of such a model analysis is given in \cite{Barata:2023jgd}.} whose unitary evolution mimics the phase information that can be reconstructed from $E^2 I^2$ analyses of UPC and EIC data.

\section*{Acknowledgements}
We would like to thank Adrian Dumitru, Heikki Mantysaari, Farid Salazar, Bjoern Schenke, Chun Shen, Wangmei Zha, Wenbin Zhao, and Jian Zhou for valuable discussions. Special thanks go to Spencer Klein for his deep insights on the topic. R. V. would like to thank Selim Jochim and Maciej Ga{\l}ka for discussions on cold atom HBT. These were facilitated by the DFG Collaborative Research Center SFB 1225 (ISOQUANT) and the EMMI Rapid Reaction Task Force workshop ``Deciphering many-body dynamics in mesoscopic quantum gases" at Heidelberg University. H.D thanks the EIC Theory Institute at Brookhaven National Laboratory for its hospitality during early stages of this work.

R.V's research is supported by the U.S. Department of Energy, Office of Science, under contract DE-SC0012704 and within the framework of the SURGE Topical Theory Collaboration. He acknowledges partial support from an LDRD at BNL. R.V's work on quantum information science is supported by the U.S. Department of Energy, Office of Science, National Quantum Information Science Research Centers, Co-design Center for Quantum Advantage (C2QA) under contract number DE-SC0012704. He was also supported at Stony Brook by the Simons Foundation as a co-PI under Award number 994318 (Simons Collaboration on Confinement and QCD Strings). The work of XZB
is supported in part by the U.S. DOE Office of Science under contract Nos. DE-FG02-89ER40531, DE-SC0012704, DE-FG02-10ER41666, and DE-AC02-98CH108. The work of JDB is supported in part by the U.S. Department of Energy, Office of Science, Office of Nuclear Physics, under contract number DE-SC0024189 as part of the Early Career Research Program. H.D is supported by NSF Nuclear Theory grant 2208387, and partially by US-Israel Binational Science Foundation grant 2022132 and the U.S. Department of Energy, Office of Nuclear
Physics through contract DE-SC0020081. The work of Z.T is supported by the U.S. Department of Energy under Award DE-SC0012704 and the Laboratory Directed Research and Development LDRD-23-050 project.

\appendix
\titleformat{\section}[block]{\centering\normalfont\bfseries}{Appendix \thesection.}{1em}{}

\section{Hanbury-Brown--Twiss effect}
\label{Appendix:A}
The Hanbury-Brown--Twiss(HBT) effect refers to the intensity correlations induced by extended sources of light \cite{Baym:1997ce}. Its essence can be demonstrated by the following model illustrated in Fig.~\ref{fig:HBT}. Imagine we have an extended source of single frequency ($\omega$) photons. We also have two detectors $\alpha$ and $\beta$, which are far away from the source. In what follows, 1 and 2 correspond to arbitrary locations on the source; to perform the measurement, we will  average the amplitude over the entire source.
\begin{figure}[H]
    \centering
\includegraphics[width=0.5\linewidth]{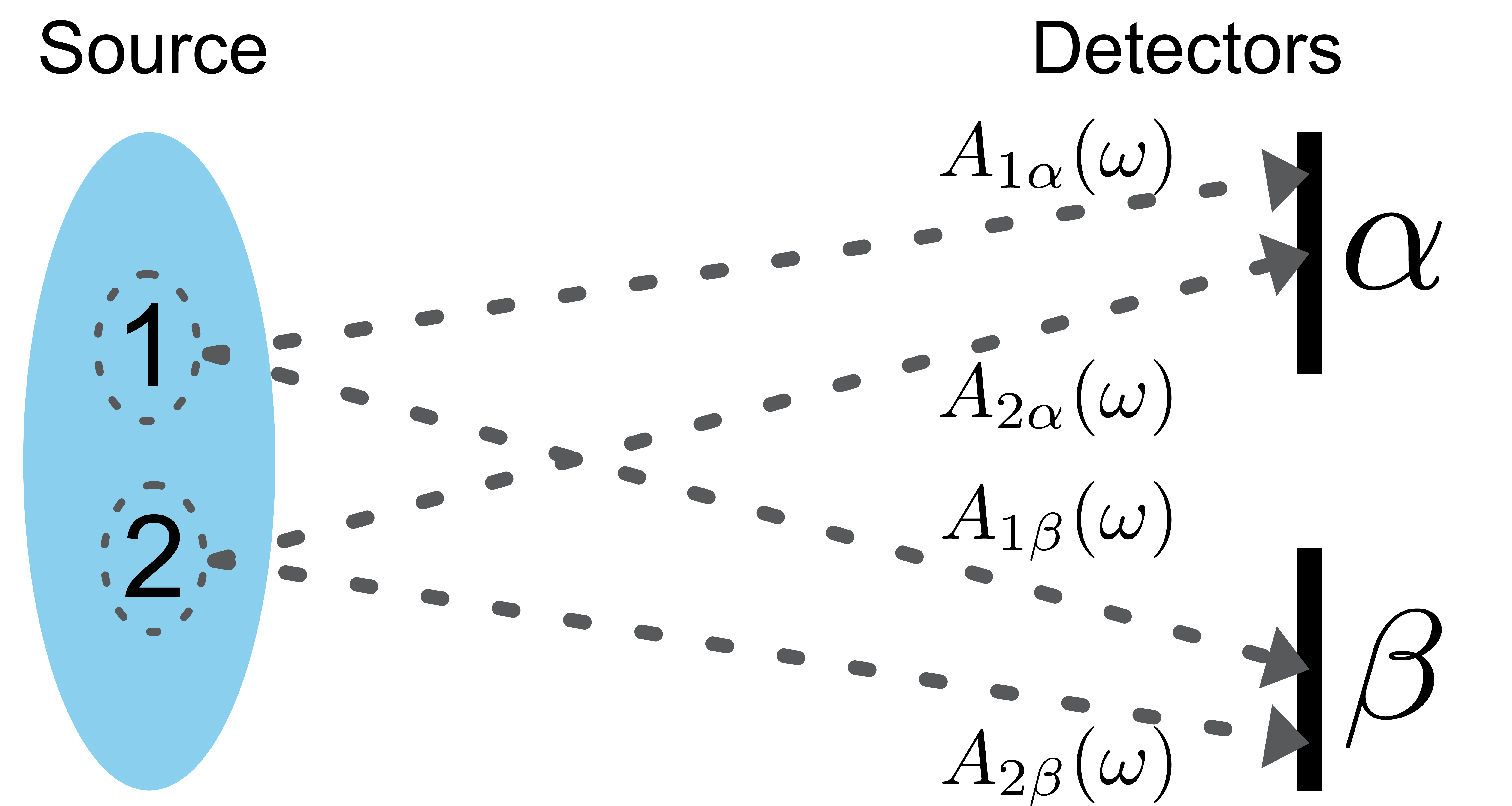}
    \caption{An illustration of Hanbury-Brown--Twiss intensity inteferometry. The regions 1 and 2 correspond to uncorrelated regions in the source, and $\alpha$ and $\beta$ are detectors that measure in coincidence light with identical frequency $\omega$. $A_{1(2)\alpha(\beta)}$ represents the amplitude for a photon from source 1(2) to be measured by detector $\alpha(\beta)$. }
    \label{fig:HBT}
\end{figure}
Schematically, we can write the instantaneous amplitude at the detectors as,
\begin{equation}
    \begin{split}
        A_{\alpha}&= A_{1\alpha}+ A_{2\alpha} \\
        A_{\beta}&= A_{1\beta}+ A_{2\beta}\,.
    \end{split}
\end{equation} 
Because of the stochastic nature of the source, the phase associated with the location of the amplitude will fluctuate from point to point and vanish when we average over all locations:
\begin{equation}
    \langle A_{1\alpha}\rangle \propto \int_0^{2\pi} \frac{d\theta_1}{2\pi} e^{i\theta_1}=0 \,,
\end{equation}
where $\theta_1$ is the initial phase of the amplitude. It is straightforward to see the following quantities will vanish as well,
\begin{equation}
        \langle A_{2\alpha}\rangle =0\,\,;\,\,
          \langle A_{1\alpha}A^*_{2\alpha}\rangle =0\,.
\end{equation}
The last result is because the average over two points will factorize, $\langle A_{1\alpha}A^*_{2\alpha}\rangle =\langle A_{1\alpha}\rangle \langle A^*_{2\alpha}\rangle=0$. This means we cannot observe any interference in an individual detector:
\begin{equation}
\langle I_\alpha\rangle=|A_{1\alpha}|^2+|A_{2\alpha}|^2 \,\,;\,\,
\langle I_\beta\rangle=|A_{1\beta}|^2+|A_{2\beta}|^2 \,.
\end{equation}

One must therefore look to intensity correlations to recover an interference pattern. This is given by
\begin{equation}
\langle I_\alpha I_\beta\rangle-   \langle I_\alpha \rangle \langle I_\beta\rangle 
    = \langle A_{1\alpha}A^*_{2\alpha}A^*_{1\beta}A_{2\beta} + (A_{1\alpha}A^*_{2\alpha}A^*_{1\beta}A_{2\beta})^* \rangle 
    =2 \Re{\langle A_{1\alpha}A^*_{2\alpha}A^*_{1\beta}A_{2\beta} \rangle}\,.
\end{equation}
Because $A_{1\alpha}$ and $A^*_{1\beta}$ come from the same point, their random initial phase will cancel. However on account of the intensity correlation, we will recover the correlation between point 1 and 2, allowing one to extract information about the target. 

We can see how this works by taking the amplitudes to be plane waves,
\begin{equation}
A_{1\alpha} \propto e^{i \v k  \cdot  \v r_{1\alpha}}\,\, ;\,\,
    A_{1\beta} \propto e^{i\v k \cdot \v r_{1\beta}}\,.
\end{equation}
We take the distance between the detectors to be d, the size of the source to be R, and the distance from the detector to the source to be L. 
Further, as was the case for the original astronomical distances motivating HBT, $L \gg R \gg d$.
We can then write,
\begin{equation}
 \Re{\langle A_{1\alpha}A^*_{2\alpha}A^*_{1\beta}A_{2\beta} \rangle} 
        = \cos(\v k \cdot ( \v r_{1\alpha}- \v r_{1\beta})-  \v k \cdot (\v r_{2\alpha}- \v r_{2\beta})) 
        \approx  \cos[(\vec k_\alpha - \vec k_\beta)\cdot (\vec r_1-\vec r_2)]\,,
\end{equation}
where $\vec k_\alpha$ is the momentum of the photon that reached detector $\alpha$. We see that for fixed momentum difference, the interference pattern is sensitive to the size of the source. 

One can simply reformulate the entire discussion in the language of quantum mechanics. Imagine we have a quantum state spanned by the Hilbert space of the two detectors,
\begin{equation}
\label{eq:state-phi}
    \begin{split}
        |\phi\rangle= \Big( A_{1\alpha} A_{2\beta} +A_{2\alpha} A_{1\beta} \Big) |\omega^\alpha,\omega^\beta\rangle \,,
    \end{split}
\end{equation}
where $|\omega^\alpha \rangle $ refers to the state vector for a photon of frequency $\omega$ that reached the $\alpha$ detector. It is easy to check that,
\begin{equation}
\label{eq:phi}
    \langle \phi | \phi \rangle - |A_{1\alpha} |^2|A_{2\beta}|^2 -|A_{2\alpha} |^2|A_{1\beta}|^2 
    =A_{1\alpha} A_{2\beta} A^*_{2\alpha} A^*_{1\beta}+A^*_{1\alpha} A^*_{2\beta} A_{2\alpha} A_{1\beta}\,.
\end{equation}
We see that this reproduces the interference pattern observed by HBT albeit the uncorrelated piece is different due to the exclusion of the situation when both photons comes from the same point. (This is because $\langle I_\alpha\rangle \langle I_\beta\rangle$ contains terms such as $|A_{1\alpha}|^2|A_{1\beta}|^2$.)

However if the two photons have different frequencies, the correct state would be
\begin{equation}
    \begin{split}
        |\psi\rangle=  A_{1\alpha} A_{2\beta}|\omega_1 \rangle^{\alpha} \otimes |\omega_2\rangle^{\beta} + A_{2\alpha} A_{1\beta} |\omega_2 \rangle^{\alpha} \otimes |\omega_1\rangle^{\beta}\,.
        \label{Eq:omega12}
    \end{split}
\end{equation} 
It is straightforward to see that {\color{blue} the }interference term will be killed by the fact that the two final states are orthogonal to each other. To recover the interference pattern for photons with different frequencies, we will need entanglement enhanced intensity interferometry, as discussed in the following appendix.

\section{Entanglement Enhanced Intensity Interferometry}
\label{Appendix:B}

\begin{figure}[H]
    \centering
\includegraphics[width=0.5\linewidth]{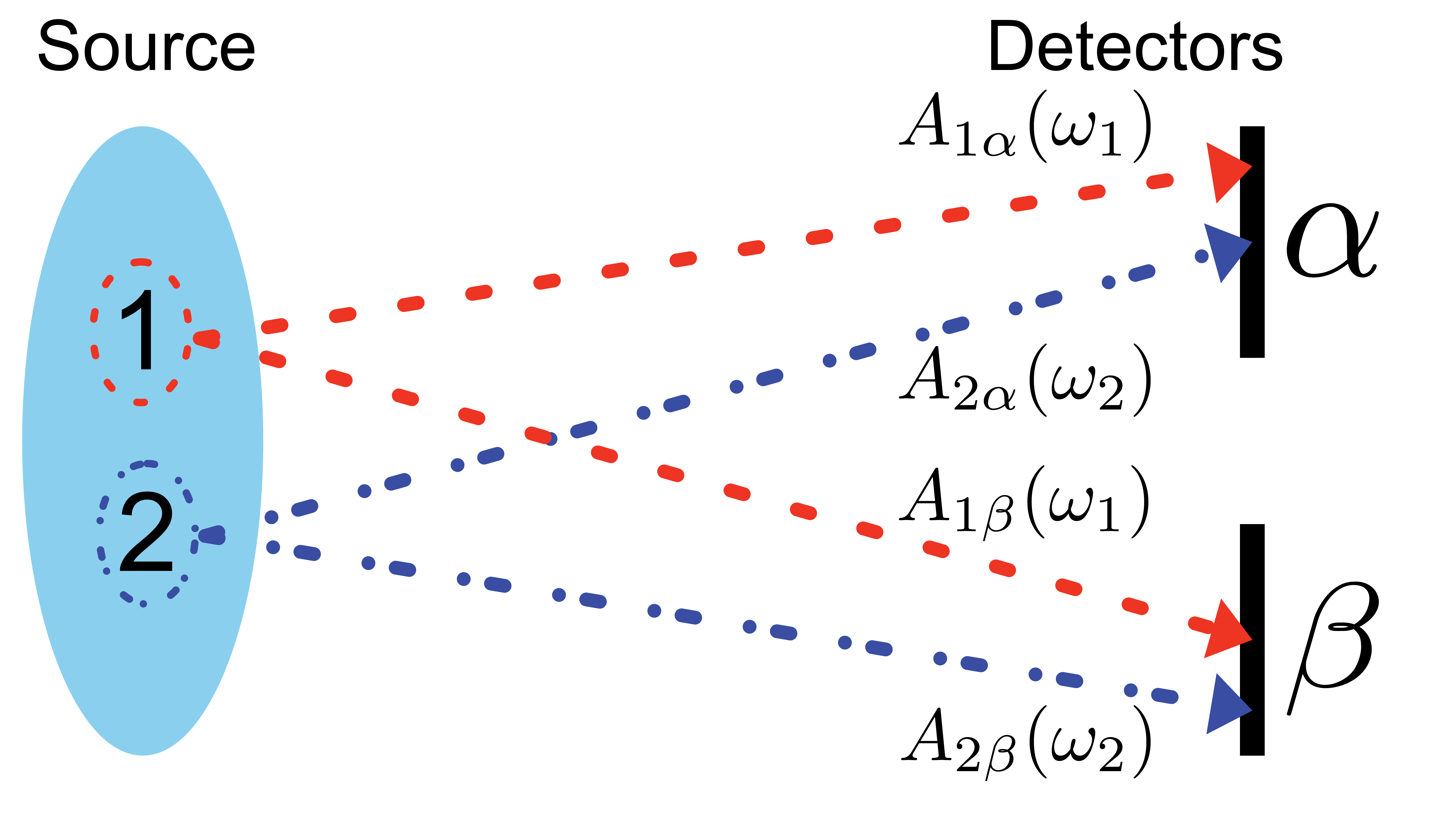}
    \caption{Coincidence measurement of light of different frequencies emitted by source 1 (red) and source 2 (blue). Without $E^2 I^2$, the HBT signal is absent.}
    \label{fig:HBT}
\end{figure}

We briefly review here the Cotler-Wilczek $E^2 I^2$ process and write it into a form that is easy to map to our problem.  The starting point is Eq.~\eqref{Eq:omega12}.
   \begin{equation}
    \begin{split}
        |\psi\rangle=  A_{1\alpha} A_{2\beta}|\omega_1 \rangle^{\alpha} \otimes |\omega_2\rangle^{\beta} + A_{2\alpha} A_{1\beta} |\omega_2 \rangle^{\alpha} \otimes |\omega_1\rangle^{\beta}\,.
    \end{split}
\end{equation} 
The goal would be to recover the interference pattern between amplitude $A_{1\alpha} A_{2\beta}$ and $A_{2\alpha} A_{1\beta}$ that would be absent in HBT. Cotler and Wilczek \cite{cotler_entanglement_2015} proposed to apply a unitary transformation $U$ on the state $|\psi\rangle$ so that $U(|\omega_1 \rangle^{\alpha} \otimes |\omega_2\rangle^{\beta})$ and $U(|\omega_2 \rangle^{\alpha} \otimes |\omega_1\rangle^{\beta} )$ are no longer orthogonal to each other. They then use a filter $\hat \Pi$ to project out the same state $| \phi \rangle$ as we had previously in Eq.~\eqref{eq:state-phi}. If the projection has the same amplitude, 
\begin{equation}
    \begin{split}
        \langle \phi | \hat \Pi U(|\omega_1 \rangle^{\alpha} \otimes |\omega_2\rangle^{\beta}) =    \langle \phi | \hat \Pi U(|\omega_2 \rangle^{\alpha} \otimes |\omega_1\rangle^{\beta} )\,,
    \end{split}
\end{equation}
the operator $\hat\Pi \,U$ does not change the relative normalization between two different outcomes, and their effect can be removed by normalizing the state. Namely,
\begin{equation}
    \begin{split}
        |\psi \rangle \rightarrow  (A_{1\alpha} A_{2\beta} + A_{2\alpha} A_{1\beta} )| \phi\rangle\,,
    \end{split}
\end{equation}
which then makes clear how to recover the interference term. To demonstrate the procedure, we can choose $U$ such that,
\begin{equation}
        \begin{split}
            U |\omega_1\rangle&= \cos(\theta)|\omega_1\rangle+\sin(\theta)e^{i\omega_0}|\omega_2\rangle\,,\\
            U |\omega_2\rangle&= \sin(\theta)e^{-i\omega_0}|\omega_1\rangle+\cos(\theta)|\omega_2\rangle\,,
        \end{split}
\end{equation}
and the filter,
\begin{equation}
    \Pi=|\omega_1\rangle^\alpha \langle \omega_1|^\alpha \otimes |\omega_1\rangle^\beta \langle \omega_1|^\beta\,.
\end{equation}
This filter projects out $|\omega_1^\alpha\rangle \otimes |\omega_1^\alpha\rangle $. One can think of this process from another perspective by applying the unitary operator U on to the filter operator. We  express this procedure in the form,
\begin{equation}
    U^\dagger \Pi U \rightarrow = |\tilde\psi \rangle \langle \tilde \psi | \,,
\end{equation}
such that $\langle \tilde \psi |$ has nontrivial overlap with both $|\omega_1 \rangle^{\alpha} \otimes |\omega_2\rangle^{\beta}$and $|\omega_2 \rangle^{\alpha} \otimes |\omega_1\rangle^{\beta}$. In our case of resonant vector meson decay of $\pi^\pm$, this is the state of $|\pi^+ \pi^-,\v p_1, \v p_2\rangle$. We still need to require that such an operation does not affect the relative norm of two distinct outcomes. Otherwise, it is difficult to extract the interference of the \textit{original} amplitudes. In general, it is difficult to control and system specific. For example, in the case of photon discussed above the angle applied to $|\omega_1\rangle$ should be the same as the angle applied to $|\omega_2\rangle$.

\newpage
\bibliography{main}

\end{document}